\documentclass[aps,prd,showpacs,showkeys,superscriptaddress,nofootinbib]{revtex4-2}

\usepackage{epsfig,graphicx,amsfonts,amsmath,amssymb,bbm}
\usepackage{bbold}
\usepackage{xcolor}
\colorlet{BLUE}{blue}
\usepackage{tikz}
\usepackage{cancel}
\usepackage{float}

\RequirePackage{mathrsfs}
\usepackage{color}
\usepackage{epsf}
\usepackage{epstopdf}

\def\eq#1{{Eq.~(\ref{#1})}}
\newcommand{\Le}{\left(}
\newcommand{\Ra}{\right)}

\newcommand{\beq}{\begin{equation}}
	\newcommand{\eeq}{\end{equation}}
\newcommand{\beqar}{\begin{eqnarray}}
	\newcommand{\eeqar}{\end{eqnarray}}
\newcommand{\D}{\partial}

\begin{document}

\title{Modeling an internal structure of a black hole using a thermodynamic quasi-particle model}

\author{S.~Bondarenko }
\email{sergeyb@ariel.ac.il}
\author{D.~Cheskis}
\email{dimach@ariel.ac.il}
\author{R.~Singh}
\email{raghvendra@ariel.ac.il}

\affiliation{Ariel University, Ariel 4070000, Israel}


\begin{abstract}
{We develop an effective thermodynamic model for a black-hole interior composed of scalar quasiparticles. The interior is represented by two regions: a dense core and a surrounding crust, whose properties are controlled by the quasiparticle kinetics. In the core, quasiparticles are assumed to have vanishing classical kinetic energy, so the total core energy is dominated by a potential-energy functional $U(N)$ that depends only on the quasiparticle number $N$. As a consequence, the appropriate intensive variable governing the core thermodynamics is an inverse-temperature--like parameter $\beta$, introduced as the thermodynamic conjugate to $U$; it replaces the usual kinetic temperature $T$ in the core equations of state and can drive the core pressure and energy density negative in the relevant regime. Different core states are further characterized by the mean occupation number $\eta$. In the crust, quasiparticles remain trapped at finite kinetic temperature, and the no-escape condition is implemented via a truncation of the phase-space integrals, yielding an explicit analytic coupling between thermodynamics and gravity. The resulting framework provides a unified quasiparticle description of core and crust, clarifies the thermodynamic origin of negative pressure/energy in the interior, and provides an effective thermodynamic setting for exploring how semiclassical or microscopic resolutions of the singularity problem might be constrained.} 
\end{abstract}
	
	\maketitle

\section{Introduction}\label{Intro}

{Black holes occupy a central position in our understanding of both gravity and quantum many-body physics. Since Bekenstein’s identification of black hole entropy with a quarter of the horizon area and Hawking’s derivation of thermal radiation \cite{Bekenstein1973,Hawking1975}, they have been regarded as thermodynamic objects with a large number of microscopic degrees of freedom, yet the nature of those degrees of freedom and the structure of the interior remain poorly understood. This has motivated a variety of approaches that recast black hole physics in language closer to statistical and condensed-matter systems. Examples include fluid-like effective descriptions of horizons and near-horizon regions \cite{ThornePriceMacdonald1986,BhattacharyyaEtAl2008}, as well as proposals in which the classical interior is replaced by many-body states such as fuzzballs \cite{Mathur2005} or gravastars / gravitational condensate stars \cite{MazurMottola2001,MazurMottola2004}. In parallel, analogue- and emergent-gravity models treat horizons as collective phenomena in underlying media, from acoustic and other analogue black holes \cite{Unruh1981,BarceloLiberatiVisser2005} to Bose–Einstein–condensate-inspired configurations \cite{GarayEtAl2000,Lahav:2010, Steinhauer:2016} and graviton-condensate or quantum-$N$-portrait pictures of black holes \cite{DvaliGomez2011,DvaliGomez2013}. Related considerations at very high densities in neutron-star and quark-matter physics \cite{OzelFreire2016,AlfordSchmittRajagopalSchaefer2008} also support viewing ultra-compact objects as strongly correlated quantum fluids with nonstandard equations of state. Taken together, these developments suggest treating black holes as thermodynamic many-body systems with an effective interior medium, which is the perspective we adopt in this work.}

{Within this thermodynamic and many-body viewpoint, the classical 	extcolor{blue}{thermodynamic behavior} of a black hole as it appears to an external observer is by now well established \cite{Beken1,Beken2,Hawk1,Hawk2,Hawk3,Strom1,Frol1,Frol2,Frol3,Page1,Page2,Page3}. However, a derivation of these thermodynamic properties directly from a concrete microscopic model of the interior is still lacking. In particular, one would like to obtain the entropy, temperature, and related quantities from the dynamics and statistics of internal degrees of freedom, and to understand how the macroscopic thermodynamic 	extcolor{blue}{behavior emerges} from the interior structure. Clarifying this connection between the black hole interior and the external thermodynamic properties is therefore an important open problem. In the present work we do not address this derivation; instead, we construct and analyse a simple effective interior model (with a core–crust structure) and its intrinsic thermodynamics, which may serve as a starting point for future studies of this connection.}

{As a first step toward such a microscopic picture, we adopt this effective-medium perspective and construct an explicit interior model. Rather than deriving the full interior geometry within classical general relativity, we treat the black hole interior as a quasiparticle medium with a core–crust structure: a maximally packed, effectively zero-momentum core surrounded by a finite-temperature, gravitationally trapped crust. Both subsystems are analysed using thermodynamic and kinetic tools familiar from condensed matter and statistical physics. Our aim is to identify a simple, internally consistent interior model that exhibits well-defined internal thermodynamic behaviour and allows one to track how these quantities depend on the assumed structure of the interior.}

{Before specifying the microscopic content in detail, it is useful to clarify how gravity would enter in a more complete treatment and how it is handled here. In a fully relativistic description one would introduce coordinates inside the horizon and define a metric there. For a spherically symmetric interior, a convenient choice is}
\beq\label{IntM1}
ds^2\,=\,d\tau^{2}\,-\,e^{\gamma(\tau,r)}dr^{2}\,-\,\mu^{2}(\tau,r)\,d^{2}\Omega\,,
\eeq
{with time and radial coordinates $\tau$ and $r$, see \cite{Land}. One would then have to solve the corresponding Einstein equations with an energy–momentum tensor derived from the proposed interior model \cite{Nov1,Nov2}, thereby obtaining a self-consistent, dynamical {spacetime}. A fully dynamical analysis of this kind, which promotes our effective stress tensor to the source of the metric, lies outside the scope of the present paper and will be addressed elsewhere.}

{Motivated by the desire to isolate the thermodynamic content of the model, we therefore adopt a highly simplified effective description of gravity. Rather than deriving a nonstatic, spherically symmetric metric, we simulate the black hole gravitational field by imposing a no-escape condition on the quasiparticles in the interior, implemented as a restriction on their momentum space in the crust: trajectories that would classically escape to the exterior are excluded. In this set-up, the gravitational field enters as an external parameter in the calculations rather than as the solution of Einstein’s equations. Concretely, we define all thermodynamic functions of interest as depending only on the radial coordinate $r$ in flat {spacetime} and analyse static, $r$-dependent profiles in Minkowski {spacetime}.}

{Correspondingly, the present work focuses on quasi-equilibrium (quasi-static) states of the system rather than on a full nonequilibrium evolution. We assume that, even in the dynamical problem, the time dependence can be effectively encoded through $r = r(\tau)$. In this formulation, the continuous evolution is replaced by a sequence of snapshots of the system at different values of external control parameters. By relating these instantaneous configurations, we aim to capture the main stages of the evolution and to identify robust thermodynamic properties of the proposed core–crust interior.}

 The next thing we need to introduce is the internal structure of the interior. The interior we construct is based on an interpretation of the 
idea about an infinite density limit of the matter and borrowing an example of a neutron star's interior, see \cite{Nov1,Neut1,Neut2,Neut3,Neut4} and references therein.
An assumption is that a black hole is a factory of matter-energy packing, or more precisely, it is an object that allows for the concentration of maximum matter (energy) in a minimal volume, i.e., an object with the highest possible density achieved in its core. In turn, this highest possible density does not mean infinite density if we stay in the framework of finite physics; it is assumed that we have a large and finite density of matter inside the core. 
For the purpose of describing such a state, we introduce quasiparticles on the stage. For the sake of simplicity, we consider one kind of scalar quasiparticle from which the core is built. The request of the universal, maximal, and finite density is clarified, then through an assumption that there is the same maximally dense packing of these effective quasiparticles for all types of black holes. This state of the quasiparticles means that they have only potential energy and no kinetic energy inside the core.
In this simple picture, the black hole has only two subsystems of quasiparticles. 
The first one is a spherical core of quasiparticles with maximal possible density
and with radius $r_{0}$ smaller than the Schwarzschild radius. The second is a much thinner crust where the quasiparticles are produced 
from the ordinary matter
at some critical conditions, further, the quasiparticles are condensing into the core under the gravity action.
Further, we do not really discuss the internal structure of the quasiparticles nor the way of their production, they are introduced in order to clarify the special state of the core.
It is merely assumed that the quasiparticles have maximally possible binding of their components, i.e., they are very heavy and can dissociate into constituents at some critical values of parameters inside the crust.
This whole quasiparticles system we can consider as an adiabatically isolated system of two different parts when external fluxes of energy-matter are absent, otherwise, the system is open and in general requires a non-equilibrium approach for its description.

 The notion of binding of ordinary matter constituents with consequent creation of some composite microscopic objects inside high-density stars is well known, of course, see for example \cite{Neut2,Neut3,Neut4} and references therein. Also, in \cite{Wil1,Wil2,Wil3,Wil4} papers, for example, a new state of matter built from quarks which interact through attractive channels and paired as bosons in antisymmetric color combinations was studied, see also \cite{Nozi}. 
We do not discuss in the paper the constituents of the introduced quasiparticles, as already mentioned. We only assume that the quasiparticles' appearance is a result of attractive interactions between the fundamental constituents of matter and that the binding energy of the constituents combined into the quasiparticle is the maximum we can achieve. Consequently, the internal structure of the
quasiparticle must satisfy only this request, we do not really need to know much about the internal structure then. Namely, the number or kind of the quasiparticle's constituents are not really important,
see also \cite{Nozi} where a similar problem was discussed in the case of the boson liquid. 
	extcolor{blue}{The next issue we need to clarify is the quantum statistics of the quasiparticles.} Usually, dense systems are considered at low temperatures in the form of Fermi surface fermions with a condensate of bound composite bosons as excited states of the system, see \cite{Land1} for example. Nevertheless, this picture is doubtful when we talk about hot systems 
with rapidly growing density at not-so-low temperatures.
Indeed, the Fermi surface appears in the non-dense systems with weak repulsive interactions at very low temperatures; the large density limit in such systems is usually applied after the initial cooling of the system. 
In the proposed scheme, we assume that we have no low temperature regime from the very beginning, but instead a very high density of matter in the interior's bulk
from the very initial moments of the evolution. 
It means that when the temperature is not low but density and gravity are large, the pairing of the elementary fermions into the heavier composite bosons can happen before the fermions fulfill the corresponding Fermi sphere. 
If the density continues to grow and the gravity does as well, then the fermion constituents of the quasiparticle will not dissociate but, {on} the contrary, the binding of fermions will be a preferable process from the point of view of the density of the packing of matter inside the black hole. 
When the density of the core is very high, even some possible cooling of the state will not change the situation because there will not be enough phase space for the bound states dissociation. In this scheme, further growth of the density leads to the condensation of the Bose quasiparticles in the physical space. Namely, the quasiparticles fulfill the geometrical (physical) volume of the core, creating a new matter state; their zero momentum in this state is a manifestation of the maximally achieved matter density.

 In general, the proposed physics is an analogue of the Fermi particle condensation, with the difference that now we have a real condensation of the Bose particles in the
coordinate space. Following this analogy, we conclude again that these core's quasiparticles have zero momentum and that the ordinary core's temperature is very low and not really changing due to the core's evolution. Important, that for the high density limit we discuss, the zero kinetic energy of the quasiparticles is not a consequence of the low temperature but a consequence of the high density and the condensation of the quasiparticles in the coordinate space. The consequent decrease of the core's temperature is the result of the further condensation of the particles in an adiabatically isolated system. Interesting to note that this possible process is somehow inverse with respect to the low-temperature normal fermion gas, where particles condense because of a decrease in the temperature. 
Therefore, assuming that the limits of very low temperatures and very high densities are 	extcolor{blue}{not directly interchangeable}, we consider further a system of quasiparticles as subject to Bose-Einstein statistics. The quasiparticles we consider are in the highest density packing state, 	extcolor{blue}{i.e.,} these composite bosons in the condensate will fill all possible coordinate quantum states in the system. It can be a finite number, the bosons are composite and massive, but still, the number must be large when we talk about the coordinate phase space. 
Correspondingly, due to the Pauli principle, in the adiabatically closed finite volume system, the dissociation of the bound state into the fermions will be impossible. Indeed, there will not be enough different quantum states for the separated, unbinding fermions. Moreover, such a dissociation will decrease the energy density of the core, we will have less bond energy and fewer particles per unit volume. This is a situation opposite to what we are looking for if we talk about a matter and energy-packed maximally dense.

 There is an additional important difference between the proposed thermodynamic system from the usual condensation of a normal gas of fermions. 
The usual fermion gas in an open system with positive inward energy flux will be heated, and the condensation will disappear. In the proposed two-layer systems of the quasiparticles, the heating will lead only to the
change of the temperature and dynamics of a crust, the physical condensation of the quasiparticles into the core will continue and will become even {stronger}\footnote{A funny analogue of this process is the formation of scale in a kettle when boiling water. With a continuous flow of hot water into the kettle, the scale grows faster, it will not disappear even if there is no water in the kettle and the temperature decreases.}. So, for the core, we discuss a dynamical system of zero-momentum quasiparticles with only potential energy changing. Therefore, the thermodynamics of this system does not depend on the regular temperature, denoted further as $T$, but there is a need to introduce another type of temperature. This new inverse temperature
parameter, denoted as $\beta$, is related to the dual dynamics of the core, whereas only the potential energy of the quasiparticles is changing. 

 The crust layer in this system of interest is a thin layer above the core where the quasiparticles are not condensed but created and appear after a possible release from the core. In the true equilibrium state, i.e., without external fluxes, we propose that the system of core and crust is in equilibrium similar to the black body thermal equilibrium. In our case, the black body is a crust system, and the massive quasiparticles play the role of photons. Consequently, there is a {thermodynamic equilibrium process} of the quasiparticles creation and absorption in the crust. The process consists of creation and dissociation of quasiparticles in the crust, plus quasiparticles' absorption and release into/from the core. So, without the external fluxes, the system is in equilibrium, and the temperature (regular one) of the crust and the state of the core remain the same. The external fluxes, in turn, change the values of the thermodynamic parameters of the crust, i.e., its temperature and chemical potential. This non-equilibrium change leads to a corresponding change of the core's state, and a new quasi-equilibrium state appears. 
As we mentioned already, the overall process is not an equilibrium one, of course, so the different quasi-equilibrium states will have different parameters of the crust and core. The first main parameter of the crust we have in this situation is a temperature $T$. The chemical potential of the crust in an equilibrium state is always zero; the change of the crust's chemical potential occurs only due to the external sources of matter and/or energy. An additional very important crust's property we introduce in the framework is a limitation of the momentum of the quasiparticles in the crust in all corresponding phase space integrals. Indeed, we do not consider the geometry of the interior and do not directly implement a no-escape behavior of the matter after the crossing of the Schwarzschild radius.
So, this limitation plays the role of a no-escape condition for the quasiparticles in the crust; its implementation prohibits quasiparticles from leaving the crust
in the outward direction. For this new definition of the phase integral limits, a new parameter that characterizes the crust state arises. This parameter is a ratio of the crust's gravity value to the regular temperature of the crust, it can be small or large depending on the interplay of values of the gravity, produced mainly by the core, and the temperature of the crust, which depends on the inward and outward fluxes in general.

 Therefore, the manuscript is organized as follows. In Section~\ref{SetU}, we introduce the basic definitions of the approach needed for the formulation of the corresponding {thermodynamic} framework. In Section~\ref{Core}, we construct and discuss the thermodynamics of the core, whereas in Sections~\ref{Crust}--\ref{Crust1} {the structure and thermodynamic characteristics} of the crust are explored. In Section~\ref{QES}, we clarify the structure of the whole core--crust system and consider stages of black-hole evolution in the model. In Section~\ref{Res}, we bring together the main results of the proposed effective model, and the last Section~\ref{Res1} {concludes} the paper and proposes further possible {developments} of the approach. In Appendixes~\ref{AppA}--\ref{AppB}, we present calculations whose results are used in the other sections.

{\noindent\textit{Scope of the framework.} The present work is a quasi-static, effective thermodynamic treatment of a trapped quasiparticle medium in the black-hole interior. Gravity is incorporated operationally via (i) an $r$-dependent potential entering the crust integrals and (ii) a no-escape (trapping) constraint that bounds the momentum window. A self-consistent dynamical metric solving the Einstein--matter system with the corresponding stress tensor is deferred to future work; here we isolate the thermodynamic content of the core, the crust, and their interface conditions.}

\section{{Problem setup}}\label{SetU}

 The per-particle energy we introduce for the interior has the following form:
\beq\label{PSU1}
\epsilon_{q}\,=\,-\,\Theta(r_{0}\,-\,r)\,U_{0} \,+\,\Theta(r_{cr}\,-\,r)\,\Theta(r\,-\,r_{0})\,\varepsilon_{p}(p,r)\,.
\eeq
here the $r_{0}$ is a radius of the core and the $r_{cr}$ is a radius of the crust layer above the core, both of them are smaller than the Schwarzschild radius.
The $\varepsilon_{p}(p,r)\,$ can be taken as an {excited state} above the ground one given by $U_{0}$. The $U_{0}$ we define as
\beq\label{PSU101}
U_{0}\,=\,U_{p\, 0}\,-\,m\,,
\eeq
with $U_{p\, 0}$ as a potential energy of the particle inside the core due {to} gravity and
$m$ as an effective mass of the quasiparticle in $c\,=\,1$ units, see also further.  {We take $U_{p,0}$ to be constant and equal for all quasiparticles in the core. This is a leading-order approximation appropriate for a nearly homogeneous, maximally packed core state.}\footnote{{A slightly more quantitative justification follows from the core occupation formula, Eq.~\eqref{Chap10}. If one allows for a weak spatial dependence of the interior potential, $U_p(r)=U_{p,0}+\delta U(r)$ with $|\beta\,\delta U(r)|\ll 1$, then the integrand may be expanded around the volume-averaged value $U_{p,0}$. The linear term in $\delta U(r)$ vanishes after integration, while the first nontrivial correction is quadratic, of order $\beta^2\langle\delta U^2\rangle$. Thus, the approximation $U_{p,0}=\mathrm{const}$ is self-consistent to leading order for a nearly homogeneous core. A more complete treatment would require a self-consistent determination of $U_p(r)$ from the Einstein equations together with a microscopic construction of the quasiparticle state.}}


We also note that, in general, the sign of the $U_{0}$ can be arbitrary
but we consider a case of only positive $U_{0}$ {which underlines the no-escape condition} for the quasiparticle.

 Talking about the phase space integrals of the core and crust, we will face an integral in which we will need to determine the upper limit of integration over momenta. Namely, 
for the core, for example, there is the following integral, which by definition can be defined as a phase-space density (density of quantum states):
\beq\label{PSU2}
\rho_{p}\,=\,\int^{p_{max}}\frac{d^3 p}{(2\pi \hbar)^{3}}\,=\,\frac{V_{p}}{(2\pi \hbar)^{3}}\,=\,\frac{4\pi p_{max}^{3}}{3 h^{3}}\,.
\eeq
We use the standard momentum--space measure $d^3p/(2\pi\hbar)^3$. The symbol $V_p$ denotes the momentum--space volume of the ball of radius $p_{\max}$.
The maximal admissible value of $p_{\max}$ in these (and similar) integrals depends on the subsystem under consideration.. For the crust, it can be defined through a no-escape condition, see \eq{Cru5} in the Section ~\ref{Crust} further. In the case of the core system of the quasiparticles, the maximal momentum can be defined through an uncertainty principle as
\beq\label{PSU2001}
p_{max}\,=\,h / \lambda_{p}\,
\eeq
for some $\lambda_{p}$.
In the {thermodynamic} system we discuss,
the $\lambda_{p}$, in turn, could be defined as a thermal de Broglie wavelength for some given $T_{min}$:
\beq\label{PSU3}
\lambda_{p}\,=\,\frac{h}{\sqrt{2\pi m T_{min}}}\,\rightarrow\,p_{max}\,=\,\sqrt{2\pi m T_{min}}\,,
\eeq
where the possible value of the $T_{min}$ {cannot be directly determined} in the present set-up.
Assuming in any case that the quasiparticles have a zero kinetic energy, we see that the $T_{min}$ 
corresponds to the quantum motion of the quasiparticles and must be small.
So we obtain for the phase-space density:
\beq\label{PSU4}
\rho_{p}\,=\,\frac{4\pi}{3}\Le \frac{2\pi m T_{min}}{h^{2}}\Ra^{3/2}\,,
\eeq
here and further $c\,=\,1$ is taken in the corresponding expressions.
The usual particle density of the system's core is defined as usual, with $V(r)$ denoting the physical volume of a sphere of radius $r$:
\beq\label{PSU5}
\rho_{N}\,=\,\frac{N}{V(r)}\,=\,\frac{3}{4\pi}\,\frac{N}{r^{3}}\,.
\eeq

It is important to notice that the $\rho_{N}$ has some finite (constant) value at the given $T_{min}$ temperature 
for any value of $N$, even when we discuss an asymptotically large density limit. Indeed, 
the highest possible density of the core's matter is accounted by the request of the maximally dense packing of the quasiparticles inside the core
, i.e. by an absence of the classical motion of the quasiparticles. Therefore, 
the \eq{PSU5} expression for the density is about the maximally possible packing of any
number of the quasiparticles which fulfill corresponding finite volume. The only difference in the
densities therefore can be due the $\lambda_{p}$ uncertainty in quasiparticle position from \eq{PSU2001} which depends on the $T_{min}$.
The size of this volume, or the value of radius in \eq{PSU5}, consequently is not 
arbitrary but {directly depends} on the way of packing of the quasiparticles at some temperature, i.e. 
\beq\label{PSU5001}
r\,=\,r(N)\,=\,\gamma(T_{min})\,N^{1/3}\,;\,\,\,\,\rho_{N}\,=\,\frac{N}{V(r)}\,=\,\frac{3}{4\pi}\,\gamma^{-3}(T_{min})\,
\eeq
with direct dependence of $\gamma$ on $T_{min}$ and $T_{min}\,=\,T_{min}(N)$ as well, see \eq{Sub001} further.
The value of the density therefore depends on the quasiparticles structure, physics of their organization into the new state, temperature of the core
but not directly on the number of the quasiparticles in the core.

 Thereby, further, in different expressions we obtain answers {whose final form depends} on the interplay between these two densities. 
We notice in turn that the ratio
\beq\label{PSU7001}
\eta\,=\,\frac{\rho_{N}}{\rho_{P}}\,=\,\Le \frac{3}{4\pi}\Ra^{2}\,\frac{h^{3}}{\Le 2\pi m T \gamma^{2}(T)\Ra^{3/2}}\,
\eeq
has a simple meaning, it is a mean occupation number of our system of quasiparticles in the core. {In the present framework, $\eta$ should be viewed as an internal state parameter of the effective model, distinguishing dilute and highly occupied regimes of the condensed core.} At $T\,\rightarrow\,0$
the number is large, in an opposite limit the number is small.
Namely, when 
\beq\label{PSU7002}
\eta\,\ll\,1
\eeq
the system is similar to the system of classical, discrete and distinguishable particles (i.e. Boltzmann statistic particles). 
Whereas, in turn, we have
\beq\label{PSU7003}
\eta\,\gg\,1\,,
\eeq
we face a system which describes by Bose-Einstein statistics (quantum particles description). 
Now we can rewrite the \eq{PSU7002} as
\beq\label{PSU7004}
\gamma^{2}(T_{min})\,T_{min}\,\gg\,\frac{h^{2}}{2\pi m}\Le \frac{3}{4\pi}\Ra^{4/3}\,
\eeq
and \eq{PSU7003} as
\beq\label{PSU7005}
\gamma^{2}(T_{min})\,T_{min}\,\ll\,\frac{h^{2}}{2\pi m}\Le \frac{3}{4\pi}\Ra^{4/3}\,
\eeq
correspondingly. 

 Next we remind, that in the proposed core's thermodynamic system
there is an another notion of temperature we need to introduce. As mentioned already, it is denoted as $\beta$ ($\beta$ as inverse temperature) in the expressions.
Technically the reason for a new temperature is simple. We talk about a dual thermodynamics with zero kinetic energy quasiparticles with only a potential energy present
in the energy expression. In this case, the regular temperature of our system, which is a measure of the kinetic energy, must be zero and {cannot be used for the thermodynamic description} of such dual systems. 
So, for that dual or "inverse" thermodynamics, we introduce a new inverse " temperature" parameter with an arbitrary sign,
negative or positive. The sign, value, and physical meaning of this parameter can be explored through the usual definition of the inverse temperature parameter in a derivative of an internal energy with 
respect to entropy; this task is performed in the corresponding Sections.
Concerning the processes in the crust layer, we also note {that within the model there exists, albeit not directly,} a notion of "lattice energy", see \cite{Latt} for example. 
Namely, we assume that there is a process of the quasiparticle "construction" (or dissociation) where some amount of energy 
\beq\label{PSU8}
\delta Q_{con}\,=\,-\Delta\,T_{bath}\,=\,-\,\delta Q_{bath}
\eeq
of the heat bath is taken for the binding of the 
usual matter's constituents into the quasiparticle at some value of the gravity field. In the proposed model this binding energy is coded in the quasiparticle effective mass $m$, see \eq{PSU101}, the mass can be very large of course.
In the opposite direction, the dissociation of the quasiparticle into its constituent parts occurs with the release of the same amount of energy $\delta Q_{con}$. 
In any case the quasiparticles creation and/or dissociation depends, perhaps, on the bath's temperature and local value of the gravity.

\section{Core's thermodynamic profile}\label{Core}

 As we noticed already, we consider the following $\epsilon_{q}$ energy profile of the Bose quasi-particles inside the core:
\beq\label{Intr1}
\epsilon_{q}\,=\,-\,\Theta(r_{0}\,-\,r)\,U_{0} \,
\eeq
the $r_{0}$ here , as mentioned, is a radius of a black hole core built from the quasi-particles. 
In order to introducing the Gibbs distribution function of the quasi-particles we have to define a kind of temperature in the core whereas the particles have no kinetic energy, see discussion in the previous Section.
Generally speaking, therefore, the usual definition of the temperature is not well stated 
and for the internal energy of the core determined as
\beq\label{Intr4}
U\,=\,-\,U_{0} \,N_{0}\,=\,-\,\rho_{N_{0}}\,V\,U_{0} \,
\eeq
see \eq{PSU5}, we introduce an inverse temperature $\beta$ parameter
\beq\label{Intr5}
\frac{1}{T}\,\rightarrow\,\beta\,,\,
\eeq

{The parameter $\beta$ acts as an inverse-temperature--like control for potential-energy exchange in the core, where classical kinetic motion is frozen. It is \emph{distinct} from the usual kinetic temperature $T$ used for the crust and should not be conflated with it: in core-sector formulas $1/\beta$ replaces $1/T$, whereas in the crust we keep the standard temperature $T$ throughout.}
{We also emphasize that the subscript $0$ will denote quantities associated with the core, while the subscript $1$ will denote quantities associated with the crust.}
Correspondingly, 
we have for the Bose distribution of the black hole interior's particles the following expression:
\beq\label{Chap6}
dN\,=\,g\,/\,\Le e^{-\,\beta\,\Le U_{0} \,+\,\mu\Ra\,\Theta(r_{0}\,-\,r)\,+\,\Theta(r_{cr}\,-\,r)\,\Theta(r\,-\,r_{0})\,
\Le \varepsilon_{p}(p,r)-\,\mu\Ra /T}\,-\,1 \Ra\,
\frac{d^3 x\,d^3 p}{ (2\pi \hbar)^{3}}\,
\eeq
with $g$ as a number of internal degrees of freedom of the quasiparticle, we consider it as a scalar one taking $g\,=\,1$.
The whole number of particles in the system we obtain integrating the \eq{Chap6} over the phase space:
\beq\label{Chap7}
N\,=\,\frac{1}{(2\pi \hbar)^{3}}\,\int\,d^3 p\,\int_{0}^{r_{0}}\,d^3 r\,\frac{1}{e^{-\,\beta\,\Le U_{0} \,+\,\mu\Ra}\,-\,1}\,+\,
\frac{1}{(2\pi \hbar)^{3}}\,\int\,d^3 p\,\int_{r_{0}}^{r_{cr}}\,d^3 r\,\frac{1}{e^{\Le \varepsilon_{p}(p,r)-\,\mu\Ra /T}\,-\,1}\,=\,N_{0}\,+\,N_{1}\,,
\eeq
further in this Section we consider the only first term of the expression.
Defining in the first term an integration over the momentum through the \eq{PSU2}-\eq{PSU4} regularization,
we obtain:
\beq\label{Chap10}
 N_{0}\,=\,\frac{4\pi\, r_{0}^{3}\,\rho_{p}}{3}\,\frac{1}{e^{-\,\beta\,\Le U_{0} \,+\,\mu\Ra}\,-\,1}\,.
\eeq
We see that in the obtained expression the $r_{0}$ radius of the core is 
an analogue of Fermi momentum of the usual Fermi surface in momentum space at low temperatures.
Namely, for
\beq\label{Chap101}
\mu\,=\,-\,U_{0} \,-\,\frac{1}{\beta}\,\ln\Le 2\Ra\,,
\eeq
we have
\beq\label{Chap10101}
N_{0}\,=\,\rho_{p}\,V(r_{0})
\eeq
and
\beq\label{Chap11}
p_{F}\,=\,\pi^{2/3}\,\hbar\,\Le \frac{N}{V}\Ra^{1/3}\,\,\longleftrightarrow \,r_{0}\,=\,\Le\frac{N_{0}}{\rho_{p}}\Ra^{1/3}\,\Le \frac{3}{4\pi}\,\Ra^{1/3}\,=\,
N^{1/3}_{0}\,\Le \frac{3}{4\pi}\,\Ra^{2/3}\,\frac{h }{\sqrt{2\pi m T_{min}}}\,\,,
\eeq
see below and corresponding \eq{PSU7004}-\eq{PSU7005}.
The analogue is even more precise if we will put attention that in the model the particles inside the core are condensed in the sense that they are not moving classically with the
quasiparticle classical motion allowed only at $r\,>\,r_{0}$. 
The \eq{Chap10} expression can be considered as a definition of the chemical potential of the core. Introducing core's particles densities by \eq{PSU5},
we obtain:
\beq\label{Chap13}
\mu\,=\,-\,U_{0} \,-\,\frac{1}{\beta}\,\ln\Le 1\,+\,\frac{\rho_{p}}{\rho_{N_{0}}}\,\Ra\,,\,\,\,\,\,-\infty\,<\beta\,<\,\infty\,.
\eeq
There are different possible limits of \eq{Chap13} expression we have to explore now.
In the case when 
\beq\label{Chap14}
\frac{\rho_{N_{0}} }{\rho_{p}}\,=\,\eta_{0}\,\gg\,1
\eeq
the answer we obtain is simple:
\beq\label{Chap141}
\mu\,\approx\,-\,U_{0} \,-\,\frac{1}{\beta}\,\frac{1}{\eta_{0}}\,
\eeq
In an opposite case when
\beq\label{Chap142}
\eta_{0}\,\ll\,1
\eeq
the expression we obtain is the following one:
\beq\label{Chap15}
\mu\,\approx\,-\,U_{0} \,+\,\frac{1}{\beta}\,\ln \eta_{0}\,,\,\,\,\,\,-\infty\,<\beta\,<\,\infty\,.
\eeq
An another limit discussed above is given by \eq{Chap101} with $\rho_{p}\,=\,\rho_{N_{0}}$ . We have then
\beq\label{Chap18}
\mu\,=\,-\,U_{0} \,-\,\frac{1}{\beta}\,\ln\Le 2\Ra\,,
\eeq
the potential depends only on $\beta$ in this case.

 Introducing the grand thermodynamic potential for the core we can further explore the core's {thermodynamic properties}:
\beq\label{CPR1}
\Omega\,=\,\frac{1}{\beta}\,\int\,\frac{d^3 x\,d^3 p}{ (2\pi \hbar)^{3}}\,\ln\Le 1\,-\,e^{\beta\,\Le U_{0} \,+\,\mu\Ra}\Ra\,.
\eeq
Rewriting the \eq{Chap10} as
\beq\label{CPR2}
1\,-\,e^{\beta\,\Le U_{0} \,+\,\mu\Ra}\,=\,\frac{\rho_{p}}{\rho_{N_{0}}}\,e^{\beta\,\Le U_{0} \,+\,\mu\Ra}\,
\eeq
and accounting \eq{Chap13} expression, we obtain:
\beq\label{CPR3}
\Omega\,=\,\frac{1}{\beta}\,\rho_{p}\,V(r_{0})\,\Le \ln\Le \frac{\rho_{p}}{\rho_{N_{0}}}\Ra\,+\,\beta\Le U_{0} \,+\,\mu\Ra\Ra\,=\,
\frac{1}{\beta}\,\rho_{p}\,V(r_{0})\,\Le \ln\Le \frac{\rho_{p}}{\rho_{N_{0}}}\Ra\,-\,\ln\Le 1\,+\,\frac{\rho_{p}}{\rho_{N_{0}}}\Ra\Ra\,.
\eeq
Consequently we calculate the entropy of the core's system using its definition:
\beq\label{CPR4}
S\,=\,\beta^{2}\,\Le \frac{\D \Omega}{\D\beta}\Ra_{V,\mu}
\eeq
obtaining
\beq\label{CPR5}
S\,=\,-\,N_{0}\,\frac{\rho_{p}}{\rho_{N_{0}}}\,\ln\Le\frac{\rho_{p}}{\rho_{N_{0}}}\,\Ra\,-\,\beta\,N_{0}\,\Le 1\,+\,\frac{\rho_{p}}{\rho_{N_{0}}}\Ra\,
\Le U_{0} \,+\,\mu \Ra\,.
\eeq
Inserting in \eq{CPR5} the \eq{Chap13} expression we in turn obtain the per-particle entropy of the system:
\beq\label{CPR6}
S/N_{0}\,=\,-\,\frac{\rho_{p}}{\rho_{N_{0}}}\,\ln\Le\frac{\rho_{p}}{\rho_{N_{0}}}\,\Ra\,+\,\Le 1\,+\,\frac{\rho_{p}}{\rho_{N_{0}}}\Ra\,
\ln\Le 1\,+\,\frac{\rho_{p}}{\rho_{N_{0}}}\,\Ra\,,
\eeq
we see that the entropy does not depend on $\beta$, i.e. requirement of the positivity of the entropy value do not disable negative sign of the $\beta$. 
Again, there are different limits we have here.
When 
\beq\label{CPR7}
\eta_{0}\,\gg\,1
\eeq
we have:
\beq\label{CPR61}
S\,\approx\,\rho_{p} V\,\Le 1\,+\,\ln \eta_{0}\,\Ra\,
\eeq
with per particle entropy defined as
\beq\label{CPR9}
S/N_{0}\,\approx\,\frac{1}{\eta_{0}}\,\Le 1\,+\,\ln \eta_{0}\,\Ra\,.
\eeq
In the opposite limit
\beq\label{CPR701}
\eta_{0}\,\ll\,1
\eeq
we obtain correspondingly:
\beq\label{CPR8}
S/N_{0}\,\approx\,1\,-\,\ln\eta_{0}\,=\,1\,+\,\ln\Le \rho_{p} V\Ra\,-\,\ln N_{0}\,.
\eeq
Finally, for the $\eta_{0}\,=\,1$ limit we have:
\beq\label{CPR10}
S/N_{0}\,=\,2\ln 2\,.
\eeq
for any value of the $\beta$ and $N_{0}$.

 {The next thermodynamic quantity of interest} is a pressure in the core. We have simply:
\beq\label{CPR11}
P\,=\,-\,\frac{\Omega}{V(r_{0})}\,=\,-\,
\frac{1}{\beta}\,\rho_{p}\,\Le \ln\Le \frac{\rho_{p}}{\rho_{N_{0}}}\Ra\,-\,\ln\Le 1\,+\,\frac{\rho_{p}}{\rho_{N_{0}}}\Ra\Ra\,=\,
\frac{1}{\beta}\,\rho_{p}\,\Le \ln\Le \frac{\rho_{N_{0}}}{\rho_{p}}\Ra\,+\,\ln\Le 1\,+\,\frac{\rho_{p}}{\rho_{N_{0}}}\Ra\Ra\,.
\eeq
The sign of $\beta$ {now matters} in the answer, different signs of $\beta$ correspond to different directions of the pressure inside the core and
now we need to define the $\beta$. At the following  {limit}
\beq\label{CPR1101}
\eta_{0}\,\gg\,1\,,
\eeq
we use \eq{CPR9} in the following form
\beq\label{CPR12}
\frac{\D S}{\D N_{0}}\,=\,\frac{1}{\eta_{0}}\,\Le 1\,+\,\ln \eta_{0}\,\Ra\,,
\eeq
where in the calculations we used the fact that the $\eta_{0}$ at the first order approximation depends on the slowly changing $T_{min}$ temperature but not on the $N_{0}$,
see \eq{PSU7001}.
Correspondingly, using the internal energy \eq{Intr4} expression, we can define:
\beq\label{CPR14}
\frac{1}{\beta}\,=\,\Le \frac{\D U}{\D S}\Ra_{V}\,=\,\Le\frac{\D U }{\D N_{0} }\Ra_{V}\,\Le\frac{\D N_{0} }{\D S }\Ra_{V}\,
=\,-\,U_{0}\,\Le\frac{\D N_{0} }{\D S }\Ra_{V}\,=\,-\,\eta_{0}\,U_{0}/\Le 1\,+\,\ln \eta_{0}\Ra\,,
\eeq
i.e. in general it could be positive or negative, in our definition the sign of $U_{0}$ is always positive, see \eq{PSU101}.
So, in this limit
we rewrite the \eq{CPR11} pressure as
\beq\label{CPR15}
P\,=\,-\,\rho_{N_{0}}\,U_{0}\,\ln\eta_{0}/\Le 1\,+\,\ln \eta_{0}\Ra\,,
\eeq
the pressure's sign as well depends on the sign of $U_{0}$ in general.
An another interesting consequence of the \eq{CPR14} expression is the \eq{Chap141} answer for the chemical potential at this limit, we have there:
\beq\label{CPR16}
\mu\,\approx\,-\,U_{0} \,-\,\frac{1}{\beta}\,\frac{\rho_{p}}{\rho_{N_{0}}}\,=\,-U_{0}\,\ln\eta_{0}/\Le 1\,+\,\ln \eta_{0}\Ra\,.
\eeq
Calculating the heat capacity of the core, we have two different possible answers related to the two different temperatures. Namely, rewriting the 
\eq{Intr4} expression as 
\beq\label{CPR17}
U\,=\,-U_{0}\frac{S \eta_{0}}{1\,+\,\ln \eta_{0}}\,=\,\frac{S}{\beta}\,,
\eeq
we can define two heat capacities with respect to the two introduced temperatures. The first one is defined with respect to the usual \eq{PSU4} temperature,
we have correspondingly:
\beq\label{CPR18}
C_{VT}\,=\,\Le \frac{\D U}{\D T}\Ra_{V}\,=\,\frac{V}{\beta}\,\frac{\D \rho_{p}}{\D T}\,+\,V\,\rho_{p}\,\frac{\D (1/\beta)}{\D T}\,=\,
0\,,
\eeq
see \eq{CPR14} definition. 
The second heat capacity we can introduce is defined with respect to the $\beta$, we have then:
\beq\label{CPR1801}
C_{V\beta}\,=\,-\,\beta^{2}\,\Le \frac{\D U}{\D \beta}\Ra_{V}\,=\,S\,=\,\frac{N_{0}}{\eta_{0}}\,\Le 1\,+\,\ln \eta_{0}\,\Ra\,,
\eeq
see \eq{PSU5001} and \eq{PSU3}.
This heat capacity is positive and small at $\eta_{0}\,\gg\,1\,$.

 Considering the case when 
\beq\label{CPR19}
\eta_{0}\,\ll\,1
\eeq
we obtain in turn by differentiation of \eq{CPR8}:
\beq\label{20}
\Le\frac{\D S}{\D N_{0}}\Ra_{V}\,=\,-\ln N_{0}\,+\,\ln\Le \rho_{p}V \Ra\,=\,-\,\ln \eta_{0}\,=\,\ln \Le 1/\eta_{0}\Ra\,.
\eeq
Correspondingly we have:
\beq\label{CPR21}
\frac{1}{\beta}\,=\,\Le \frac{\D U}{\D S}\Ra_{V}\,=\,-\,U_{0}\,\Le\frac{\D N_{0} }{\D S }\Ra_{V}\,=\,U_{0}/\ln \eta_{0}\,;
\eeq
that provides the following chemical potential
\beq\label{CPR22}
\mu\,\approx\,-\,U_{0} \,+\,\frac{1}{\beta}\,\ln \eta_{0}\,=\,0\,
\eeq
and pressure
\beq\label{CPR2201}
P\,=\,\frac{\rho_{p}}{\beta}\,\ln\Le 1\,+\,\eta_{0}\Ra\,=\,\rho_{p}\,U_{0}\,\frac{\ln\Le 1\,+\,\eta_{0}\Ra}{\ln \eta_{0}}\,\approx\,-\,\rho_{p}\,
U_{0}\,\frac{\eta_{0}}{\ln\Le 1/ \eta_{0}\Ra}\,.
\eeq
Using the following rewritten expression for the internal energy
\beq\label{CPR220101}
U\,=\,-\,\frac{N_{0}}{\beta}\,\ln\eta_{0}\,,
\eeq
we obtain for the following answers
\beq\label{CPR23}
C_{V T}\,=\,0\,;\,\,\,C_{V\beta}\,=\,-\,\beta^{2}\,\Le \frac{\D U}{\D N_{0}}\Ra_{V}\,\Le \frac{\D N_{0}}{\D \beta}\Ra_{V}\,=\,N_{0}\,\ln\Le 1/\eta_{0}\Ra\,
\eeq
for the heat capacities values.

 For the last value of the mean occupation number
\beq\label{CPR24}
\eta_{0}\,=\,1\,
\eeq
we obtain the following values of the {thermodynamic quantities}.
The number of particles we have in this state and {corresponding energy} are:
\beq\label{CPR25}
N_{0}\,=\,\frac{S}{2\ln 2}\,,\,\,\,\,U\,=\,-\,\frac{S U_{0}}{2\ln 2}\,;\,
\eeq
it provides correspondingly
\beq\label{CPR26}
\frac{1}{\beta}\,=\,\Le \frac{\D U}{\D S}\Ra_{V}\,=\,-\,\frac{U_{0}}{2\ln 2}\,
\eeq
and
\beq\label{CPR27}
P\,=\,\frac{\rho_{p}}{\beta}\,\ln 2\,=\,-\,\frac{U_{0} \rho_{p}}{2}\,=\,-\,\frac{U_{0} \rho_{N_{0}}}{2}\,,
\eeq
see \eq{CPR11} expression for $\eta_{0}\,=\,1\,$.
For the chemical potential we obtain the following answer:
\beq\label{CPR28}
\mu\,=\,-\,U_{0} \,-\,\frac{1}{\beta}\,\ln\Le 2\Ra\,=\,-\,\frac{U_{0}}{2}\,,
\eeq
and
\beq\label{CPR29}
C_{V T}\,=\,0\,;\,\,\,C_{V\beta}\,=\,2\,N_{0}\,\ln 2\,.
\eeq
for the values of the heat capacities.

\section{Thermodynamic profile of the crust: averaged number and internal energy of quasiparticles}\label{Crust}

 Consider the second term of the \eq{Chap7} expression:
\beq\label{Cru1}
N_{1}\,=\,\frac{1}{(2\pi \hbar)^{3}}\,\int\,d^3 p\,\int_{r_{0}}^{r_{cr}}\,d^3 r\,\frac{1}{e^{\Le \varepsilon_{p}(p,r)-\,\mu\Ra /T}\,-\,1}\,
\eeq
with
\beq\label{Cru2}
\varepsilon_{p}(p,r)\,=\,\varepsilon_{k}(p)\,-\,U_{p}(r)\,+\,m\,;\,\,\,U_{p}(r)\,>\,0\,.
\eeq
For the changing number of quasiparticles in the crust {due to their absorption and creation}, the $N_{1}$ in a {quasi-equilibrium state} is determined by the condition of the minimum of the free energy of the system in the:
\beq\label{Cru3}
\frac{\D F}{\D N_{1}}\,=\,0\,=\,\mu\,;
\eeq
see for example in \cite{Land1}, consequently we obtain:
\beq\label{Cru4}
N_{1}\,=\,\frac{(4\pi)^2}{(2\pi \hbar)^{3}}\,
\int\limits^{p_{max}}_{p_{min}}\,p^2\,d p\,\int_{r_{0}}^{r_{cr}}\,r^2 d r\,\frac{1}{e^{\varepsilon_{p}(p,r)/T}\,-\,1}\,\approx\,
\frac{(4\pi)^2 r_{cr}^{2} d}{(2\pi \hbar)^{3}}\,
\int\limits^{p_{max}}_{p_{min}}\,\frac{p^2\,d p\,}{e^{\varepsilon_{p}(p,r_{cr})/T}\,-\,1}\,;\,\,\,d/r_{cr}\,=\,(r_{cr}\,-\,r_{0})/r_{cr}\,\ll\,1\,;
\eeq
where the per particle energy of a quasiparticle in the crust we write approximately as following:
\beq\label{Cru4001}
\varepsilon_{p}(p,r)\,\approx\,\varepsilon_{p}(p,r_{cr})\,=\,\varepsilon_{k}(p)\,-\,U_{p}(r_{cr})\,+\,m\,=\,\varepsilon_{k}(p)\,-\,\tilde{U}_{p}(r_{cr})\,.
\eeq
{A usual change of variable is performed next:}
\beq\label{Cru5}
\frac{\varepsilon_{k}(p,r_{cr})}{T}\,=\,\frac{p^2}{ 2m T}\,=\,z\,;\,\,\,p_{max}\,=\,\sqrt{2 m U_{p}(r_{cr})}\,;
\,\,\,p_{min}\,=\,\sqrt{2 m \tilde{U}_{p}(r_{cr})}\,.
\eeq
The minimal momentum is determined by definition of the potential energy of the quasiparticle from the core's radius, i.e. by $\varepsilon_{p}(p_{min},r_{cr})\,=\,0$;
the value of the maximal momentum is an implementation of the no-escape condition correspondingly.
Then we obtain
\beq\label{Cru6}
N_{1}\,=\,
\frac{(4\pi)^2 r_{cr}^{2} d \Le 2 m T\Ra^{3/2} }{2(2\pi \hbar)^{3}}\,
\int\limits^{U_{p}(r_{cr})/T}_{\tilde{U}_{p}(r_{cr})/T}\,\frac{z^{1/2}}{e^{z-\tilde{U}_{p}(r_{cr})/T}\,-\,1}\,dz\,
\eeq
with
\beq\label{Cru601}
\tilde{U}_{p}(r_{cr})\,=\,U_{p}(r_{cr})\,-\,m\,.
\eeq
The expression determines the $N_{1}$ as a function of {the given temperature $T$} of the crust in a {quasi-equilibrium state}.
The maximum momentum here is determined through a no-escape condition for the particle in the crust, the discussion on the
occupation numbers of Section ~\ref{SetU} is valid in this case as well with only $T_{min}\,\rightarrow\,U_{p}(r_{cr})$ substitution performed.
So, in general, there are the following possible asymptotic regimes for the \eq{Cru6} we have. The first when 
\beq\label{Cru7}
m / T\,\gg\,1\,;\,\,\,\,\tilde{U}_{p}(r_{cr})/T\,\gg\,1\,;
\eeq
the other with
\beq\label{Cru10}
m / T\,\ll\,1\,;\,\,\,\,\tilde{U}_{p}(r_{cr})/T\,\geq\,1\,;
\eeq
and the last
\beq\label{Cru1002}
m / T\,\ll\,1\,;\,\,\,\,\tilde{U}_{p}(r_{cr})/T\,\ll\,1\,.
\eeq
The final answer for the integral depends, as underlined, also on the sign of $\tilde{U}_{p}(r_{cr})$ determined by \eq{Cru601} and on the value of the $\tilde{U}_{p}(r_{cr})/T$ ratio. In our framework
we consider only the
\beq\label{Cru1001}
\tilde{U}_{p}(r_{cr})\,=\,U_{p}(r_{cr})\,-\,m\,>\,0\,
\eeq
case for the quantity, defining it as an additional no-escape condition for the quasiparticles in the crust.

 In the \eq{Cru7} limit, the leading order contribution we obtain is written in \eq{A3003}, see Appendix~\ref{AppA}. We have:
\beq\label {Cru10021}
N_{1}\,=\,\frac{2\pi }{h^{3}}\,V \Le 2 m \tilde{U}_{p}\Ra^{3/2}\,\frac{T}{\tilde{U}_{p}}\,\ln\Le T/\mathcal{E}_{0}\Ra\,
\eeq
Now, fixing the number of quasiparticles $N_{1}$ in the crust and
defining the phase space density and usual density 
\beq\label {Cru1003}
\rho_{p}\,=\,\frac{4\pi}{3 h^{3}}\,\Le 2 m \tilde{U}_{p}\Ra^{3/2}\,;\,\,\,\rho_{N_{1}}\,=\,\frac{N_{1}}{V}\,;
\eeq
we obtain that the \eq{Cru10021} can be rewritten in terms of the mean occupation number of the crust:
\beq\label {Cru1004}
\eta_{1}\,=\,\frac{\rho_{N_{1}}}{\rho_{p}}\,=\,\frac{3}{2}\,\frac{T}{\tilde{U}_{p}}\,\ln\Le T/\mathcal{E}_{0}\Ra\,.
\eeq
The equality depends on the temperature and value of $\tilde{U}_{p}$ but also from the value of $\mathcal{E}_{0}$, at 
$T \rightarrow \mathcal{E}_{0}$ limit the mean occupation number is zero.
The \eq{Cru10}, in turn, provides:
\beq\label {Cru1005}
N_{1}\,=\,\frac{2\pi }{h^{3}}\,V \Le 2 m \tilde{U}_{p}\Ra^{3/2}\,\frac{T}{\tilde{U}_{p}}\,\ln\Le m/\mathcal{E}_{0}\Ra\,
\eeq
and
\beq\label {Cru10051}
\eta_{1}\,=\,\frac{\rho_{N_{1}}}{\rho_{p}}\,=\,\frac{3}{2}\,\frac{T}{\tilde{U}_{p}}\,\,\ln\Le m/\mathcal{E}_{0}\Ra\,;
\eeq
this number depends not only on the temperature and $\tilde{U}_{p}$ but also on the quasiparticle's mass $m$. At the $\mathcal{E}_{0}\rightarrow 0$ 
and large $m$ the number can be large.
The last, \eq{Cru1002} condition, is manifested through the form of \eq{A30} answer. If we take into account the only first from the leading terms of the expression 
\beq\label{Cru1006}
N_{1}\,=\,\frac{2\pi }{h^{3}}\,V \Le 2 m \tilde{U}_{p}\Ra^{3/2}\,
\Le \frac{T}{\tilde{U}_{p}}\Ra^{3/2}\,\Le \sqrt{\pi}+
\ln(\frac{T}{\tilde{U}_{p}})\,\Le\frac{\tilde{U}_{p}}{T}\Ra^{1/2}\,\Ra\,\approx\,\frac{2\pi^{3/2} }{h^{3}}\,V \Le 2 m T\Ra^{3/2}\,.
\,,
\eeq
The mean occupation number we have in this case is
\beq\label{Cru1007}
\eta_{1}\,=\,\frac{\rho_{N_{1}}}{\rho_{p}}\,=\,\frac{2}{3\sqrt{\pi}}\,\Le \frac{T}{\tilde{U}_{p}} \Ra^{3/2}\,\gg\,1\,
\eeq
due the \eq{Cru1002} condition.
 
 Next we consider an internal energy of the quasiparicles in the crust, we have:
\beq\label{Cru14}
U\,=\,\frac{1}{(2\pi \hbar)^{3}}\,\int\,d^3 p\,\int_{r_{0}}^{r_{cr}}\,d^3 r\,\frac{\varepsilon_{k}(p)\,-\,U_{p}(r)\,+\,m}{e^{\varepsilon_{p}(p,r)/T}\,-\,1}\,
\eeq
Assuming a smallness of the width $d$ of the crust, we take in the first approximation $U_{p}(r)\,\approx\,U_{p}(r_{cr})$ in the expression and write:
\beq\label{Cru15}
U\,\approx\,\frac{(4\pi)^2 r_{cr}^{2} d \Le 2 m T\Ra^{5/2} }{4m\,(2\pi \hbar)^{3}}\,\Le \frac{\tilde{U}_{p}(r_{cr})}{T}\Ra^{5/2}\,
\int\limits^{m/\tilde{U}_{p}}_{\mathcal{E}_{0}/\tilde{U}_{p}}\,\frac{\Le z\,+\,1\Ra^{3/2}}{e^{\tilde{U}_{p}(r_{cr})\,z/T}\,-\,1}\,dz\,-\,N_{1}\,\tilde{U}_{p}(r_{cr})\,,
\eeq
see Appendix~\ref{AppA} for the definition of the integral's limits.
The integral is calculated in Appendix~\ref{AppB}, we have a suitable perturbative series for the each from \eq{Cru7}-\eq{Cru1002} limits. Here we represent only leading contributions to the obtained series, see \eq{B6}, \eq{B8} and \eq{B14} in Appendix~\ref{AppB}. So we have:
\beq\label{Cru1501}
U/N_{1}\,=\,\mathcal{E}_{0}\,
\eeq
in the case of \eq{Cru7} limit with the $N_{1}$ given by \eq{Cru10021}. Correspondingly, the answer
\beq\label{Cru1502}
U/N_{1}\,=\,-\,m\,\frac{\ln\Le T/m \Ra}{\ln\Le m/\mathcal{E}_{0}\Ra}
\eeq
we have for the \eq{Cru10} limit and \eq{Cru1005} $N_{1}$. The last, \eq{Cru1002} limit in the integrals, provides
\beq\label{Cru1503}
U/N_{1}\,=\,T/2\,-\,\tilde{U}_{p}\,
\eeq
answer for the \eq{Cru1006} averaged number of the quasiparticles in the crust.
The corresponding heat capacities we have are the following therefore. For the \eq{Cru1501} expression valid for \eq{Cru7} condition and not small temperatures
we obtain:
\beq\label{Cru1504}
C_{V1}\,=\,\Le \frac{\D U}{\D T}\Ra_{V}\,=\,\frac{\mathcal{E}_{0}}{T}\,N_{1}\Le 1\,+\,\ln^{-1}\Le T/\mathcal{E}_{0} \Ra\,\Ra\,\approx\,
U/T\,,
\eeq
Introducing a temperature related to $\mathcal{E}_{0}$, see \eq{A3004}, we obtain in turn
in the $T\,\rightarrow\,\mathcal{E}_{0}\,=\,T_{min}$ limit:
\beq\label{Cru1505}
C_{V1}\,=\,\Le \frac{\D U}{\D T}\Ra_{V}\,=\,N_{1}\,+\,
\frac{2\pi }{h^{3}}\,V \Le 2 m \tilde{U}_{p}\Ra^{3/2}\,\frac{T_{min}}{\tilde{U}_{p}}\,\approx\,
\frac{2\pi }{h^{3}}\,V \Le 2 m \tilde{U}_{p}\Ra^{3/2}\,\frac{T_{min}}{\tilde{U}_{p}}\,.
\eeq
It is accounted here that $N_{1}\,\sim\,\ln\Le T/T_{min} \Ra$, we see that {we cannot neglect} the second term in this case of a very low temperature.
For the \eq{Cru1502} internal energy and \eq{Cru10} condition we have correspondingly:
\beq\label{Cru1506}
C_{V2}\,=\,\frac{U}{T}\,\Le 1\,+\,\ln^{-1}\Le T/m \Ra\Ra\,<\,0\,;
\eeq
the heat capacity is negative at this temperature.
The last expressions we have are \eq{Cru1503} and \eq{Cru1006} valid at \eq{Cru1002} limit. 
In this case {the heat capacity we obtain is} the following:
\beq\label{Cru1508}
C_{V3}\,=\,\frac{5}{2}\frac{U}{T}\,+\,\frac{\tilde{U}_{p}}{T}\,,
\eeq
i.e.,to the leading order approximation, the form of this formally high-temperature answer coincides with expression for the usual low-temperature Bose gas.

\section{Thermodynamic profile of the crust: grand {thermodynamic} potential, pressure and entropy } \label{Crust1}

 Next we introduce a grand {thermodynamic} potential for the crust, we define it as usual for the some quasiequilibrium state with constant $N_{1}$
and zero chemical potential:
\beq\label{Cru22}
\Omega\,=\,T\,\int\limits^{r_{cr}}_{r_{0}}\,d^3 r\,\int\limits^{p_{max}}_{p_{min}}\,
\frac{d^3 p}{ (2\pi \hbar)^{3}}\,\ln\Le 1\,-\,e^{\Le U_{p}(r)\,-\,\varepsilon_{k}(p)\,-\,m \Ra / T}\Ra\,.
\eeq
In the thin crust approximation $U_{p}(r)\,\approx\,U_{p}(r_{cr})$ with $r_{cr}\,-\,r_{0}\,=\,d\,\ll\,r_{0}\,$,
we perform an integration by parts with respect to the $p$ variable obtaining:
\beqar
\Omega\,&=&\,\frac{4\pi T V}{3 h^{3}}\,\Le
p_{max}^{3}\,\ln\Le 1\,-\,e^{\Le U_{p}(r)\,-\,\varepsilon_{k}(p_{max})\,-\,m \Ra / T}\Ra\,-\,
p_{min}^{3}\,\ln\Le 1\,-\,e^{\Le U_{p}(r)\,-\,\varepsilon_{k}(p_{min})\,-\,m \Ra / T}\Ra\Ra\,-\,
\nonumber \\
&-&\,
\frac{2}{3}\,\Le U\,+\,\tilde{U}_{p}\,N_{1}\Ra\,.
\label{Cru2201}
\eeqar
The values of $p_{max}$ and $p_{min}$ are defined in \eq{Cru5}, we also regularize the $p_{min}$ by $\mathcal{E}_{0}$ value and have then:
\beq
\Omega\,=\,\frac{4\pi T V}{3 h^{3}}\,\Le
\Le 2 m U_{p}(r_{cr})\Ra^{3/2}\,\ln\Le 1\,-\,e^{-\,m / T}\Ra\,-\,\Le 2 m \tilde{U}_{p}(r_{cr})\Ra^{3/2}\,\ln\Le 1\,-\,e^{-\,\mathcal{E}_{0}/ T}\Ra\Ra\,-\,
\frac{2}{3}\,\Le U\,+\,\tilde{U}_{p}\,N_{1}\Ra\,.
\label{Cru2202}
\eeq
The last term integral in the expression is calculated in Appendix ~\ref{AppB}, so we still have three different contributions correspond to the three different regimes.

 For the \eq{Cru7} limit, collecting all terms together, we obtain that to leading approximation
\beq\label{Cru23}
\Omega\,=\,-\,\frac{2}{3}\,U\,-\,\frac{4\pi T V}{3 h^{3}}\,
\Le 2 m U_{p}(r_{cr})\Ra^{3/2}\,e^{-\,m / T}\,\approx\,-\,\frac{2}{3}\,U\,
\eeq 
i.e. we have here an usual answer for the gas of {nonrelativistic} Bose particles. 
So, the leading order pressure's value we obtain now is the following one:
\beq\label{Cru2303}
P\,=\,-\,\frac{\Omega}{V}\,=\,\frac{2}{3}\,\frac{U}{V}\,=\,\frac{2}{3}\,\frac{N_{1}}{V}\,\mathcal{E}_{0}\,=\,\frac{2}{3}\,\rho_{N_{1}}\,\mathcal{E}_{0}\,.
\eeq
The answer for the entropy, in turn, has the following form:
\beqar
S\,&=&\,-\Le \frac{\D \Omega}{\D T}\Ra_{N,V}\,=\,\frac{2}{3}\,\frac{U}{T}\,\Le 1\,+\,\ln^{-1}\Le T/\mathcal{E}_{0} \Ra\Ra\,=\,
\frac{2}{3}\,\frac{N_{1} \mathcal{E}_{0}}{T}\,\Le 1\,+\,\ln^{-1}\Le T/\mathcal{E}_{0} \Ra\Ra\,=\,
\nonumber \\
&=&\,
\frac{4\pi}{3h^{3}}\,V\,\Le 2 m \tilde{U}_{p}(r_{cr})\Ra^{3/2}\,\frac{\mathcal{E}_{0}}{ \tilde{U}_{p}}\,\Le 1\,+\,\ln\Le T/\mathcal{E}_{0} \Ra\Ra\,.
\label{Cru26}
\eeqar
We see, that
the per particle entropy is decreasing in the given case when the temperature is increasing, the overall entropy is increasing in turn.

 In the case of \eq{Cru10} limit, we have correspondingly the following leading order expression for the potential:
\beq\label{Cru230201}
\Omega\,=\,-\,\frac{1}{3}\,N_{1}\,m\,\frac{\ln\Le T/m\Ra}{\ln\Le m/\mathcal{E}_{0} \Ra}\,,
\eeq
with $N_{1}$ from \eq{Cru1005} expression and $m\,>\,\mathcal{E}_{0}$ assumed as well.
In this limit the pressure we have is the following one:
\beq\label{Cru2302}
P\,=\,\frac{N_{1}}{3V}\,m\,\frac{\ln\Le T/m\Ra}{\ln\Le m/\mathcal{E}_{0} \Ra}\,
\eeq
and entropy is
\beq\label{Cru27}
S\,=\,-\Le \frac{\D \Omega}{\D T}\Ra_{N,V}\,=\,\frac{N_{1}}{3\ln\Le m/\mathcal{E}_{0} \Ra}\,\frac{m}{T}\,\Le 1\,+\,\ln\Le T/m\Ra\Ra\,,
\eeq
per particle entropy is decreasing in the given case as well when the temperature is increasing.

 For the last, \eq{Cru1002}, asymptotic regime we have:
\beq\label{Cru29}
\Omega\,=\,-\,\frac{1}{3}\,N_{1}\,T\,\Le 1\,-\,\frac{2}{\sqrt{\pi}}\Le\frac{\tilde{U}_{p}}{T}\Ra^{3/2}\ln\Le T/\mathcal{E}_{0}\Ra\,+\,
\frac{2}{\sqrt{\pi}}\Le\frac{U_{p}}{T}\Ra^{3/2}\ln\Le T/m\Ra\,\Ra\,.
\eeq	
Correspondingly we obtain:
\beq\label{Cru30}
P\,=\,\frac{N_{1}}{3V}\,T\,\Le 1\,-\,\frac{2}{\sqrt{\pi}}\Le\frac{\tilde{U}_{p}}{T}\Ra^{3/2}\ln\Le T/\mathcal{E}_{0}\Ra\,+\,
\frac{2}{\sqrt{\pi}}\Le\frac{U_{p}}{T}\Ra^{3/2}\ln\Le T/m\Ra\,\Ra\,\approx\,\frac{N_{1}}{3V}\,T\,
\eeq	
and	
\beq\label{Cru2701}
S\,=\,\frac{5}{6}\,N_{1}\,\approx\,\frac{5 U}{3T}\,,
\eeq	
see \eq{Cru1503} and with $N_{1}$ given by \eq{Cru1006}, there is a constant entropy per particle in this regime.
So, next, we can discuss the
thermodynamic properties of the whole system of the core and crust together.

\section{Quasi-equilibrium states of whole system} \label{QES}

 We considered the effective model of the interior which is not an equilibrium system in general. 
As always, we can discuss the quasi-equilibrium states of the core or crust {that are} in an equilibrium
during some short periods of time in between further changes are coming. 
Therefore, we first of all will clarify the thermodynamic parameters which characterize the quasi-equilibrium states of the parts of the system and after that 
will discuss which possible quasi-equilibrium states of the whole united system. {We emphasize that this is not a full dynamical evolution in general relativity, but rather a sequence of quasi-equilibrium states of the coupled core--crust system.}

 Let's discuss firstly a $T_{min}$ dependence of the core's density in \eq{PSU5001} expression 
and related question about the \eq{PSU7001} mean occupation number value. Namely, the question is what is colder, the core with larger volume (larger number of quasiparticles 
inside the core) or the smaller core with smaller quasiparticles number inside. An another question is about to which value of the core the small or large mean occupation numbers belong. We can begin from the \eq{PSU7001} expression of the $\eta$ taking $T_{min}\,\rightarrow\,0$ there and obtaining 
$\eta\,\rightarrow\,\infty$ answer \footnote{The $\gamma$ is growing or decreasing with grow or decrease of the $T_{min}$, i.e. there is a direct dependence of the function on the temperature.}. So, 
the \eq{PSU7003} (large occupation number) limit is about colder cores and \eq{PSU7002} limit describes the hotter ones. In this extend the result is similar to the low temperature Fermi particles condensation as already was mentioned above.
In turn, a precise value of the $T_{min}$ can be defined, perhaps, only in a microscopic theory
of the core's state creation. Nevertheless, we can use here a simple {thermodynamic observation} which clarifies the issue.
If we assume that initially the
core is built from $N_{in}$ number of quasiparticles and has some $T_{in}$ temperature then, for the adiabatic growing of the core, the temperature with larger number 
of the quasiparticles $N_{0}$ decrease as
\beq\label{Sub001}
T_{min}\sim\,T_{in}\,N_{in}/N_{0}\,
\eeq
simply due the energy conservation law. The argument holds till the quasiparticles in the core will acquire a momentum, i.e. till the core will almost evaporate. {Therefore, we conclude that}, the large occupation number characterizes the core with larger volume (larger
number of quasiparticles inside the core) but lower temperature in comparison to the core with smaller number of quasiparticles and higher
temperature. The result fits the intuitive concept and definition of the mean occupation number of course.
The transition between these two stages of the system is around $\eta\,\sim\,1$ , the number has no another special meaning. In any way, 
the $T_{\min}$ remains small {during} all stages of the core evolution as a consequence of a high density of the quasiparticles.

 The second issue {that we need to understand concerns} the way the different states of the core and crust are paired. 
We have three different regimes for the core, characterized by the
mean occupation number, and correspondingly we have three different regimes for the crust which properties depend on the crust's temperature (the regular one).
So, we have to define which state of the crust pairs with which state of the core and if this pairing is unique.
{We cannot directly discuss} the problem without a dynamical theory of the non-equilibrium state of the whole system and without the gravity included. For the sake of simplicity, we roughly assume that we have only three possible combinations of the pairing of the core and crust which describe different
stages of the whole system. Checking the \eq{Cru10021}, \eq{Cru1005} and \eq{Cru1006} expressions for the 
number of the quasiparticles in the crust we see that we have no quasiparticles in the crust at the very low temperatures when $T_{min}\,\rightarrow\,\mathcal{E}_{0}$,
see \eq{Cru10021}. We assume that this state of the crust {cannot be compatible} with the state of the core with large number of quasiparticles. Indeed, the large core means
large rate of a condensation of the quasiparticles from the crust into the core. It can take place only at some high temperatures and gravity in the crust, see \eq{PSU8} as a reminder that we need heat bath for the quasiparticles creation, the bath is due the inward fluxes of heat and/or energy.
Next related question is about what is happening if all external inward fluxes are reset. In this case the crust will cool down, the rate of the cooling as well will depend on the overall heat balance because yet due the evaporation of the core and further dissociation of the quasiparticles the some inward heat into the crust flow will be present as well. 
This evaporation of the crust is accompanied by the decrease of the core's value and {heating}  of it, i.e. by change its mean occupation number and volume. The overall evolution of the whole system will be toward some equilibrium state of the core and crust therefore, till the core will completely evaporate and the only regular matter will remain inside the
hole at some low equilibrium temperature. 

 So, the pairing of the system's parts we consider is the following. We discuss the whole system consists the following pairs: core with the low occupation number and low temperature crust, core with unit occupation number and intermediate temperature crust and finally high occupation number core with high temperature crust. Interesting to note, that from point of view of combined temperatures, there is a low-high pairing, i.e. a low temperature of the core means high temperature of the crust and vice versa.

\subsection{Core with low occupation number value and crust's low temperature regime}\label{SubS1}

 Consider the \eq{PSU7002} and \eq{Cru1002} asymptotic regime of the black hole and remind 
that generally the two phases system of the non-isolated system is not in an 
equilibrium. In any case the direction of the process evolution can be clarified by the request of an entropy growing,
at the moment of an observation of a quasi-equilibrium stage of the process it is equal to zero.
Consider now, for example, a particular process of transition of quasiparticle from the core to the crust at very low temperatures:
\beq\label{Sub1}
\delta N_{1}\,=\,-\,\delta N_{0}\,=\,\delta N\,>\,0
\eeq
with 
\beq\label{Sub1001}
\delta U_{core}\,=\,\delta U_{crust}\,=\,\delta U\,>\,0\,;\,\,\,\delta V_{1}\,=\,-\,\delta V_{0}\,=\,\delta V\,.
\eeq
For the 
quasi-equilibrium state of the overall system we write
\beq\label{Sub2}
d S\,=\,\Le \frac{1}{T_{1}}\,+\, \beta\Ra\,\delta U\,+\,\Le \frac{P_{1}}{T_{1}}\,-\,\beta\,P_{0} \Ra\,\delta V\,-\,
\Le \frac{\mu_{1}}{T_{1}}\,-\, \beta\,\mu_{0}\Ra\,\delta N\,=\,0\,.
\eeq
The chemical potentials of the both sub-systems are zero, see \eq{CPR22} answer and general request of $\mu_{1}\,=\,0$
for the crust, at low temperatures of the crust it is assumed that
a chemical potential due the external flows is negligible. 
Also, here and further we do not discuss the pressure values, without gravity account it is not really meaningful. So we assume that the equilibrium provide
$P_{1}\,=\,P_{2}\,=\,0$ when gravity effect is included. 
The immediate result of this quasi-equilibrium state expression is
the crust's temperature. In this state is given by the following identity:
\beq\label{Sub2001}
T_{1}\,=\,-\,1/\beta\,=\,U_{0}/\ln\Le 1/\eta_{0}\Ra\,,\,\,\,\eta_{0}\,\ll\,1\,;
\eeq
see \eq{CPR21}. For the low $\eta_{0}$ and known $U_{0}\,=\,U_{0}(N_{0})$ the expression connects the temperature of the crust with the number 
of the quasiparticles in the core or vice versa when the $T_{min}$ temperature is known. Any process of the quasiparticles condensation or dissociation 
in or out the core changes the value of the $U_{0}$ and corresponding temperature of the crust, the opposite is also correct of course. In this extend
the microscopic description of the process will clarify the details of the process but we always will have 
\beq\label{Sub3}
\Delta\,T \,\propto\,\Delta\,U_{0}\,.
\eeq
The expression \eq{Sub2001} will remain valid till $\eta_{0}\,\ll\,1$ will be satisfied, when the crust's temperature is growing we will pass
to the intermediate regime for the crust and mean occupation number regime for the core, the core's temperature will decrease and
we also will need to consider the external flows.
This situation, when the inward fluxes are larger than the outward one, in this system's state describes a very initial stage of growing of the "young" black hole.

 In the reverse time regime we have a stage with decreasing inward fluxes and with only outward flux present finally. 
The normal asymptotic regime in this case is
\beq\label{Sub2002}
T_{1}\,=\,-\,1/\beta\,=\,U_{0}/\ln\Le 1/\eta_{0}\Ra\,\xrightarrow[U_{0},\eta_{0}\,\rightarrow\,0]\,0\,.
\eeq
Thermodynamically the process is ending by the 
alignment of the {thermodynamic parameters} of the core and crust, the temperatures and chemical potentials will change toward zeroes. Physically it means a
full evaporation of the core with further dissociation of the quasiparticles in the crust into the regular matter. The remnant is the full evaporation of the black hole or creation of an exotic object similar to the vacuum black hole, see for example discussion in \cite{Mott}.
We propose further that the hole's
outward flux is caused by the matter inside the black hole, so the end of this stage means also an end of the outward Hawking radiation.
This stage of the system we can consider as a final stage of a black hole.

\subsection{Core with unit occupation number value and crust's intermediate temperature regimes}\label{SubS2}

 For the quasi-equilibrium processes which we consider now, this combination of the parameters as well describes two
opposite in time possible stages of the whole system. These are transition from hot to lower temperature crust and opposite, from low to higher temperature crust transition.
The quasi-equilibrium temperature of the crust in this case is provided by \eq{CPR26}:
\beq\label{Sub4}
T_{1}\,=\,-\,1/\beta\,=\,U_{0}/\Le 2\ln2\Ra\,,\,\,\,\eta_{0}\,\sim\,1\,.
\eeq
Now we put attention that the \eq{CPR28} $\mu_{0}$ chemical potential of the core is not zero anymore but negative. 
For the crust in an equilibrium the chemical potential must remain zero
nevertheless. So, we need to account now the change of the chemical potential of the crust due the external flows. We write, therefore, the
$\mu_{1}$ chemical potential
as a sum of its equilibrium part and non-equilibrium contribution coming from the external sources:
\beq\label{Sub401} 
\mu_{1}\,\rightarrow\,\mu_{1}\,+\,\mu_{ext}\,,
\eeq
where, the first term in the r.h.s remains zero of course.

 Consider firstly again the core to crust transition, i.e.core evaporation process, as in \eq{Sub1}-\eq{Sub2} defined. We will have for the quasi-equilibrium:
\beq\label{Sub402}
d S\,=\,-\,\Le \frac{\mu_{ext}}{T_{1}}\,-\, \beta\,\mu_{0}\Ra\,\delta N\,=\,\beta\,
\Le \mu_{ext}\,+\,\mu_{0}\Ra\,\delta N\,\geq\,0\,.
\eeq
For $\beta\,<\,0,\,\delta N\,>\,0$ we obtain then
\beq\label{Sub403}
\mu_{ext}\,+\,\mu_{0}\,\leq\,0
\eeq
or 
\beq\label{Sub40301}
\mu_{ext}\,\leq\,-\mu_{0}\,.
\eeq
So, this transition is allowed till the \eq{Sub403} inequality satisfied by the $\mu_{ext}$ value for the $\mu_{0}$ given by \eq{CPR28}.
Namely, there is a turning (or equilibrium) point for the $\mu_{ext}$ 
\beq\label{Sub404}
\mu_{ext.\,eq.}\,=\,|\mu_{0}|\,=\,\frac{U_{0}}{2}\,
\eeq
which separates the processes of evaporation and growing of the core at given states of the system. 
For all values of the $\mu_{ext}$ smaller than the $U_{0}/2$, we will have a transition of the quasiparticles from the core to the crust.
For the larger than that value of the $\mu_{ext}$, the core will grow, see further.
This decrease of the potential, in turn, can be provided by the dominant outward flows only and it is compatible with the discussed evaporation of the core.
The process is dynamical and non-linear of course, the $U_{0}$ value depends on the number of the quasiparticles in the core and decreases
(by absolute value) with the evaporation.

 Talking about the crust to core transition, we have to take $\delta N\,=\,-1$ in \eq{Sub402} expression. Correspondingly we will obtain:
\beq\label{Sub405}
\mu_{ext}\,+\,\mu_{0}\,\geq\,0
\eeq
 and 
\beq\label{Sub40302}
\mu_{ext}\,\geq\,-\mu_{0}\,.
\eeq
as mentioned above.
We see, that the core's grow is possible only for the large and positive value of the external chemical potential, it is possible only if we have 
a positive heat balance inside the crust through the inward external flows. Namely, the excess of the heat balance over the equilibrium one
provided by the inward flows leads to the grow of the core.

The interesting observation we have in this case is that the heat capacity of the crust in these stages is negative, see \eq{Cru1506}. The situation, {to some extent}, is similar to what we have in the
star's system equilibrium, see \cite{Nov1,Nov2}, here the analogue of the star's outward cooling radiation is a condensation of the quasiparticles into the core.
If now the overall heat balance inside the crust is positive, the negative heat capacity means that the crust is cooling. 
In the star's evolution the cooling achieves by the expansion of the 
star and consequent decrease of the matter's density that brings the star back to the equilibrium and not to the heat burst. 
In our case, the negative capacity could mean the same, i.e. it is a sign of the
growing of the crust layer above the core with consequent process of equilibrium of the crust's matter. 
When in contrary the overall energy balance is negative, we will have an the situation will be opposite and the crust's layer will decrease with 
the additional heating of the matter in the crust.

\subsection{Core with large occupation number value and crust's large temperature regime}\label{SubS3}

 Again, because the quasi-equilibrium we request, the present stage is about continuing to grow black hole or it describes an very initial stage of massive
black hole evaporation.
The quasi-equilibrium temperature in this case is 
\beq\label{Sub5}
T_{1}\,=\,-\,1/\beta\,=\,\eta_{0}\,U_{0}/\Le 1\,+\,\ln \eta_{0}\Ra\,,\,\,\,\eta_{0}\,\gg\,1\,,
\eeq
see \eq{CPR14}. The arguments of the previous Subsection are valid in this case as well. The only difference is in the value of the equilibrium chemical potential,
we have now:
\beq\label{Sub40401}
\mu_{ext.\,eq.}\,=\,|\mu_{0}|\,=\,U_{0}\,\ln\eta_{0}/\Le 1\,+\,\ln \eta_{0}\Ra\,\approx\,U_{0}\,,
\eeq
see \eq{CPR16} answer. Interesting, that in this limit of the large temperature of the crust, the quasiparticles in an equilibrium behave as usual Bose particles 
at low temperatures. The result is clear, we notice that to leading order approximation the contribution from the gravity to the {thermodynamic quantities} is absent
in that stage and
the only temperature dependent contributions remain in the expressions.

\section{Results} \label{Res}

 In this Section, we summarize the main results of the proposed model.
We begin as well from the core and will finish discussing the properties of the whole system. 

 The thermodynamics of the core we introduced is unusual {in several respects}. First of all, we directly implemented in the construction of the thermodynamical system
a limit of an asymptotically large density of the core. Reformulated in the framework of the approach it means that the core is built from an effective massive quasiparticles,
it is incompressible in the classical sense (the quasiparticles have no classical kinetic energy), their internal structure is not really important and the potential energy of all of them is the same,
at least at leading approximation. This
density of the core is a universal quantity for any black hole therefore, the density depends on the core's temperature but not directly
on the number of quasiparticles in it, see \eq{PSU5001}. This is the only place of an appearance of the regular temperature in the core's thermodynamics.
Namely, the $T$ arises there through 
a quantum uncertainty of the quasiparticle position given by the de Broglie thermal wavelength of the quasiparticle, see \eq{PSU3}. {We cannot directly determine} the value of the $T_{min}$ in the framework, see nevertheless \eq{Sub001} and discussion in the end of the Section. All other quantities, we derived for the core's system, depend on the 
$\beta$ inverse "temperature" parameter defined as usual through core's internal energy dependence on entropy, see \eq{CPR14} and further.
This parameter is not the same as a regular temperature but of course it plays a role of the regular temperature in the corresponding expressions.
Therefore, the $\beta$ can be defined as an inverse "temperature" in the system with only potential energy present, see \eq{Chap7}-\eq{Chap10}.

This result, of course, is a consequence of the new type of thermodynamics for the
Bose particles system we consider\footnote{See similar construction for the Fermi particles in \cite{FB,FB1,FB2,FB3}.}.
Technically, this "dual" thermodynamics of the core is relatively simple, see results in Section ~\ref{Core}. Every {thermodynamic} parameter,
including $\beta$, can be calculated in terms of the potential energy of the quasiparticles, its mass, temperature of the core
and $\gamma$ function in \eq{PSU5001} expression which determines the radius of the core as a function of the temperature $T_{min}$. In general, 
the potential energy depends on the other parameters of the problem through the Einstein equations, particularity it must depend on $N_{0}$ which is a number of the quasiparticles
in the core. 
The description becomes even simpler if we introduce a mean occupation number of the core in terms of its quasiparticles density and 
density of quantum states, \eq{PSU5} and \eq{PSU4}. The number characterizes the core's state, its small value describes a core of a black hole in an initial 
stage of hole's growing or very late state of the hole's evaporation.
The large value of the occupation number describes, in turn, a black hole in its later evolution stages, it can be of course the further grow or beginning of hole's evaporation, see
discussion above in Section ~\ref{QES}. The transition between these different states is similar to the criterion of the Boltzmann approximation in the description 
of an ideal gas, see \cite{Land1} and \eq{PSU7004}-\eq{PSU7005}. From this point of view, the small occupation number is about a Boltzmann, classic description of the core whereas
the large occupation {number} is about the quantum description of the core. {In this latter case}, the core represents a kind of quantum unified coherent state, namely {we cannot distinguish} the 
quasiparticles inside the core but only a number of them. This number of the quasiparticles determines the core's radius, its temperature dependence, its mass. All these parameters are accommodated all together in a single expression for the mean occupation number. It is interesting to note, that in many aspects, see \eq{Intr4} \eq{CPR15},\eq{CPR2201} and \eq{CPR27} 
answers for the energy and pressures, this new state is similar to the introduced many years ago $\mu$-vacuum state of \cite{Glin,Glin1}. The self-consistent approach
to this problem requires, of course, to solve the Einstein equations for the core using \eq{IntM1} metric for example. Perhaps, the negative values of the core's pressure and energy will change the
standard form of the solutions of these equations, see also further. Concerning other important parameters of the core we notice that the per-particle entropy of the core is decreasing 
with the core's growing as well as the chemical potential and heat capacity defined in respect to the $\beta$. 
So, in conclusion, {we emphasize that we achieved a relatively simple description} of the core of black hole in the presented model. {It requires only a few parameters} for the thermodynamical description of the {black-hole} core and reveals some unexpected features of its {thermodynamic properties} calculated
it at the different stages of the evolution. 

 The thermodynamic description of the crust, in turn, looks simpler. The quasiparticles in the crust have a kinetic energy but yet there are some unusual features in the system we accounted.
First of all, talking about the black hole interior without solving corresponding gravity equations, we have to implement a no-escape condition in the corresponding integrals. Namely, in the \eq{Cru1}
integral we {limit} the corresponding phase space. From above it is limited by a no-escape condition which restrict the possible value of the
momentum and from below by definition of the potential energy level of the quasiparticle in the crust from the core radius. It must be clear, that the both values in \eq{Cru6} depends on the value of the gravitational field in the crust which, in turn, depends on the gravity of the core which, in turn, depends on the dynamics of the quasiparticles, i.e. their creation and absorption processes. The self-consistent formulation of the problem is highly non-linear and in the full formulation must include corresponding solution of the
gravity equation and spontaneous resolution of a {thermodynamically nonequilibrium system} of quasiparticles with inward and outward fluxes included. Therefore, we considered a simplified model.
We assumed, that without external sources of matter and energy, the quasiparticles behavior in the crust is similar to the black body radiation. Namely,
the quasiparticles creation, absorption and release are an equilibrium in the full core-crust system.
The difference from the equilibrium system of photons is in that the quasiparticles are massive, the similarity is that here we also assume zero chemical potential
in an equilibrium. The next simplification we made 
is that the gravity in the crust is homogenous, i.e. it does not depends on the coordinates. Clearly, the approximation is accurate enough only when the width of the crust's layer 
is much smaller than the radius of the core. 
Nevertheless, {it cannot significantly affect the main results} of the calculations. In general, if a $r$ dependence of the potential energy in \eq{Cru4} integral is present, the integral can be {calculated perturbatively, with the ratio of the given radii serving as a small parameter}. 
In this extend, the presented result is simply a leading term in this kind of a compliment perturbative answer. 
Next, as we underlined already, the crust's system in general is not in an equilibrium. During the paper we consider and analyze the quasi-equilibrium states of the system
which, by definition, are snap-shots of the system's non-equilibrium evolution. In this construction we, therefore, combine the equilibrium calculations with a non-equilibrium 
evolution using following assumptions. 
We assume that in the framework, where the non-equilibrium changes in the system due the inward and outward external sources, these external fluxes affects directly only on the 
values of the parameters present
in the quasi-equilibrium description of the system. Namely, we consider kind of a superposition of two frames. The first frame is that the core and the crust are in some quasi-equilibrium state. The second frame, suited with the first one, is that the values of the parameters of this quasi-equilibrium state are defined by the external sources at the given moment of time.
This simplifies the understanding of the overall dynamics of our system, the price that we are missing details of the dynamics. So,in this approach, the changes of the core's size and chemical potential of the crust are happening due the
external sources, without the sources the core and crust stay in an equilibrium state at constant size, chemical potential and temperature. 
In general picture in the approach we have two different asymptotic regimes for the crust with intermediate regime which describes a transition between these two. 
We can call them as low, intermediate and high temperature regimes, see \eq{Cru7}-\eq{Cru1002}. Because the other parameters we have in the approach are mass of the quasiparticles and 
value of the gravity field, the low temperature regime does not mean really low temperatures and high energy does not mean really high temperatures. 
These are only ratios of the temperature to other parameters which depend not only of th evalue of the temperature. 
Concerning the high energy regime for the crust, it is interesting to note that 
its leading order answer reproduces a very low energy behavior of the regular and free low temperature Bose gas. The similarity, of course, is not accidental but only
formal. That form of the result is reproduced because it is a leading order answer in the approximation when we neglect all other contributions, so it is not really accidental. But it is only formal because
in our calculations the full result includes  plenty {of} sub-dominant corrections where the ratio of the gravitational field and temperature present,
see Appendix~\ref{AppA} - Appendix~\ref{AppB} calculations. Having thermodynamic models of the core and crust, the next step is a description of the
united system of the black hole interior.

 There are two different limiting answers for the core exist, at small and large mean occupation numbers of the core with a transition regime
at unite value of the number. There are also two asymptotic answers for the crust as well, at low and high temperatures, with an intermediate temperature which
determined as larger than the mass of the quasiparticle. The discussion about the compatibility of the different states of the core and crust is in 
Section~\ref{QES}. The conclusion obtained there is that we have a dual system here in the following sense. If we consider an evolution from smaller to larger core, the core's regular temperature is
lowering whereas the crust temperature is growing. The same is happening with an entropy, per particle entropy of the core in respect to the $\beta$ is lowering
whereas the crust's entropy achieves some constant value at asymptotically large $T$, see \eq{Cru2701}. Concerning the chemical potential, the situation is more complicated. In the core the potential is lowering (growing by absolute values), see results of Section~\ref{Core}. In the crust the equilibrium part of the potential remains zero and then all mutual dynamics of the system is defined by the non-equilibrium part of the crust's chemical potential, i.e. by it defined by the dynamics of the external sources.

 Additional important observation we have is about the $T_{min}$ behavior, namely about it's dependence on the core's size. It was mentioned already in Section~\ref{SetU} and discussed further in Section~\ref{QES}, see \eq{Sub001}, that the larger is core (mass of the core) then the smaller $T_{min}$ we have.
This $T_{min}$ mass dependence is the same as established in many different ways for the Hawking temperature, see \cite{Haw1,Haw2}. This similarity is intriguing, but so far we have no a satisfactory explanation of this, perhaps accidental, similarity.

 Considering, approximately, the condensate of the core and the finite-temperature crust as two perfect-fluid phases with markedly different equations of state, we note the following. When those phases are embedded in a {spacetime}, the stress–energy tensor must vary smoothly across the common radius $r=r_{0}$ in order to avoid an unphysical $\delta$-function layer.  In the language of general relativity, this amounts to satisfying the Israel junction conditions and therefore requires at least pressure continuity,
\beq\label{Rag1}
P_{\mathrm{core}}(r_{0})=P_{\mathrm{crust}}(r_{0}),
\eeq
together with a suitable behavior of the energy density.  Our thermodynamic equations already allow the pressures to meet at zero, making the interface a “free” boundary in hydrostatic equilibrium .
A stricter condition—energy-density continuity,
\beq\label{Rag2}
\rho_{\mathrm{core}}(r_{0})=\rho_{\mathrm{crust}}(r_{0}),
\eeq
cannot be imposed within the present, gravity-free treatment because the core density is (large and) negative whereas the crust density is positive.  Taken at face value, this sign change would create a thin mass shell of surface density $\sigma\propto\rho_{\mathrm{crust}}-\rho_{\mathrm{core}}$ at $r_{0}$. Physically, we expect some mixing of the two phases to smear out the jump over a narrow but finite transition layer, allowing $\rho(r)$ to pass continuously through zero while $P(r)$ remains small. Introducing such a layer has no impact on the global thermodynamics discussed here and merely refines the local equation of state. This situation, {to some extent}, is similar to what we have for the degenerated Fermi gas. Namely, there instead the step function form of the distribution function at $T\rightarrow 0$, which we have as well, see \eq{PSU1} and discussion in Section ~\ref{Core}, there is some transition region around degeneracy temperature of the gas. 
In any case, since the detailed structure of the transition region cannot be resolved without solving the full Einstein field equations, we defer a quantitative analysis to our forthcoming work.

\section{Discussion and future outlook}\label{Res1}

{In this section, we extend the discussion on the main outcomes of the proposed effective model for the black-hole interior and identify the central open problems that must be addressed in order to place the framework on a fully relativistic and microscopic footing.}

{\paragraph{Summary of main results.}
The analysis presented here suggests that the asymptotically high-density interior of a black hole can be described, at an effective level, by a thermodynamic quasiparticle model with explicitly calculable macroscopic characteristics. Within this framework, one obtains a consistent set of thermodynamic variables parametrizing the interior state, from which radial profiles of core and crust quantities can be computed in a controlled approximation.}

{A central conceptual feature of the model is the emergence of an inverse-temperature--like parameter $\beta$ for the core. Since the core is treated as a coherent condensed state with suppressed classical kinetic energy, the ordinary kinetic temperature $T$ is not the natural thermodynamic variable there. Instead, $\beta$ is introduced as the thermodynamic quantity conjugate to the effective potential/binding energy of the quasiparticles. In this sense, it plays the role of an inverse temperature for the potential-energy-dominated core sector, whereas the crust remains an ordinary thermal region characterized by the standard kinetic temperature $T$.}

{A second important outcome is the natural two-phase interpretation of the interior: a high-density condensed core surrounded by a thermal crust. Under the assumption of inward and outward quasiparticle sources, the model describes the interior as a sequence of quasi-equilibrium core--crust configurations. Within this effective picture, the thermodynamic characteristics of both regions are in principle calculable, providing a useful starting point for more refined dynamical and relativistic treatments.}

{\paragraph{Relation to existing interior proposals.}
It is useful to position the present construction relative to several widely discussed scenarios for nonsingular or modified black-hole interiors.}

{\emph{(i) Gravastar / de Sitter-core models.} In gravastar-like proposals, an interior with vacuum-energy--like equation of state, often close to $p\simeq -\rho$, is postulated and then matched to an exterior Schwarzschild geometry, typically through a thin shell. Our model shares the qualitative role of negative pressure in supporting an effectively repulsive core. The difference lies in the physical origin: here the negative pressure emerges from the $\beta$-controlled thermodynamics of a maximally packed coherent phase under the trapping/no-escape constraint, rather than being imposed from the outset as a de Sitter equation of state. A precise comparison with gravastar-type constructions, however, requires a covariant stress tensor and a relativistic matching procedure, which lie beyond the scope of the present work.}

{\emph{(ii) Bose-condensate-inspired interiors.} There is clear conceptual overlap with Bose--Einstein-condensate-type pictures in which the black-hole interior is modeled as a coherent macroscopic quantum state. In our framework, the large-occupation regime and the suppression of kinetic energy place the core in the same broad class of coherent phases, while the crust retains conventional thermal behavior. A distinctive feature of the present approach is that it is formulated directly in thermodynamic terms---through $\beta$, the mean occupation number, and explicit exchange/interface conditions---rather than through a specific condensate field equation.}

{\emph{(iii) Graviton-condensate / ``$N$-portrait'' ideas.} In graviton-condensate or quantum-$N$-portrait approaches, the black hole is interpreted as a large-occupation quantum state of soft quanta, and evaporation is associated with depletion. Although the quasiparticles in our model are not assumed to be gravitons, the appearance of large occupation numbers and maximal packing suggests a possible bridge between our effective thermodynamic variables and depletion-type dynamics in a more microscopic completion.}

{\paragraph{Coupling to spacetime curvature (qualitative discussion).}
The present manuscript does not solve the Einstein equations; gravity is implemented effectively through the trapping/no-escape condition and through the potential-energy description of the interior. Nevertheless, it is useful to indicate how the derived thermodynamic profiles would enter a relativistic treatment. In spherical symmetry, one would interpret the interior as an effective medium with energy density $\rho(r)$ and, in general, anisotropic pressures $p_r(r)$ and $p_t(r)$, so that
\[
T^\mu{}_\nu=\mathrm{diag}(-\rho,p_r,p_t,p_t).
\]
A static interior configuration would then follow from the coupled Einstein--matter system, or equivalently from a TOV-type balance generalized to include anisotropy, supplemented by matching conditions at the core--crust interface. Within such a framework, negative pressure contributes nontrivially to the sourcing of curvature and may produce an effectively repulsive interior region. However, global regularity, causality, and dynamical stability can only be assessed after constructing an explicit covariant $T_{\mu\nu}$ and solving the corresponding Einstein equations. This lies beyond the scope of the present paper.}

{\paragraph{Open issues and next steps.}
The present model is intentionally effective and quasi-static. Several important questions remain open and define the main agenda for future work:}

\begin{itemize}
\item {\textbf{Covariant formulation and Einstein equations.}
The interior model developed here should be viewed as guidance for constructing a physically motivated effective source for the Einstein equations. A natural next step is to embed the thermodynamic medium into a relativistic interior metric---for example, using \eq{IntM1} or another metric compatible with the core--crust structure---and then solve the coupled system, including matching conditions and regularity requirements.}

\item {\textbf{Relativistic and nonequilibrium generalization.}
The present treatment describes quasi-equilibrium dynamics of nonrelativistic quasiparticles in flat spacetime. A more complete formulation should be developed in curved spacetime and, most likely, within a relativistic nonequilibrium framework (see, e.g., \cite{Zub,Zub1,Zub2}). Such a generalization is essential for addressing dynamical stability, causal propagation, and genuinely time-dependent processes such as accretion and evaporation.}

\item {\textbf{Microscopic origin of quasiparticles and condensation.}
A major unresolved issue is the microscopic theory behind the quasiparticles themselves: their existence, creation mechanism, internal structure, and interactions with ordinary matter. A microscopic completion is also required to justify and compute the condensation process responsible for the core state and to determine which effective degrees of freedom are physically appropriate.}

\item {\textbf{Initial core formation and the parameter $T_{in}$.}
The initial onset of core formation and the corresponding regular temperature $T_{in}$, which enters parametrically in many expressions, are not derived in the present model. Nor is their evolution computed from first principles. As indicated by \eq{Sub001}, determining $T_{in}$ and its dependence on the collapse history requires a microscopic quantum model of quasiparticle production and condensation.}

\item {\textbf{$\mu$-vacuum analogy and quantum observables.}
As noted above, the properties of the core bear some resemblance to the $\mu$-vacuum state discussed in \cite{Glin,Glin1}. It is therefore important to investigate whether this correspondence can be sharpened beyond a qualitative analogy and recast in terms of correlation functions, response properties, or other quantum observables associated with the coherent core state.}

\item {\textbf{Negative pressure/energy density and energy conditions.}
The model yields regimes in which the core pressure, and in some parameter ranges the effective energy density, become negative. In classical general relativity, many foundational results---including the Penrose--Hawking singularity theorems---assume standard energy conditions such as NEC, WEC, and SEC. If the effective core sector is interpreted as a fluid, these assumptions are not generically satisfied. Such violations are not by themselves inconsistent in an effective description, but they require a covariant analysis of regularity, causality, and dynamical stability before any claim of singularity resolution can be made.}

\item {\textbf{Singularity avoidance versus relocation of pathologies.}
A closely related issue is whether the framework genuinely leads to a geodesically complete interior or instead shifts singular behavior into a regime where the effective description itself breaks down. This question can only be settled within a self-consistent relativistic solution, together with a perturbative stability analysis.}

\item {\textbf{Heat/entropy transport and relation to Hawking radiation.}
The present paradigm describes a thermal crust coexisting with an exotic coherent core, but it does not derive the microscopic mechanism by which heat and entropy are transported from the interior to the exterior. A key goal of future work is therefore to connect the interior thermodynamics to an exterior emission law and to determine whether the framework can reproduce the universality of Hawking radiation, as well as any possible deviations in late stages of evolution.}
\end{itemize}

\appendix
\newpage
\section{ Integrals for thermodynamic quantities of the crust: number of quasiparticles}\label{AppA}
\renewcommand{\theequation}{A.\arabic{equation}}
\setcounter{equation}{0}

 We begin from the \eq{Cru6} integral rewriting it in the following form:
\beq\label{A1}
N_{1}\,=\,
\frac{2\pi }{h^{3}}\,V \Le 2 m T_{0}\Ra^{3/2}\,\Le\frac{\tilde{U}_{p}(r_{cr})}{T_{0}}\Ra^{3/2}\,
\int\limits^{U_{p}(r_{cr})/\tilde{U}_{p}(r_{cr})}_{1}\,\frac{z^{1/2}}{e^{\tilde{U}_{p}(r_{cr})\Le z-1\Ra/T_{0}}\,-\,1}\,dz\,.
\eeq
The integral we calculate we rewrite as following:
\beq\label{A2}
I\,=\,
\int\limits^{m/\tilde{U}_{p}}_{0}\,\frac{\Le z\,+\,1\Ra^{1/2}}{e^{\tilde{U}_{p}(r_{cr})\, z\,/T_{0}}\,-\,1}\,dz\,.
\eeq
Consequently we obtain:
\beqar
I\,&=&\,\int\limits^{m/\tilde{U}_{p}}_{0}\,\frac{\Le z\,+\,1\Ra^{1/2}}{e^{\tilde{U}_{p}(r_{cr})z/T_{0}}\,-\,1}\,dz\,=\,
\sum_{k=0}^{\infty}\int\limits^{m/\tilde{U}_{p}}_{0}\,\Le z\,+\,1\Ra^{1/2}\,e^{-(k+1)\tilde{U}_{p}(r_{cr})z/T_{0}}\,dz\,=\,
\nonumber \\
&=&\,
\sum_{k=0}^{\infty} e^{\frac{\tilde{U}_{p}}{T_{0}}(k+1)}\,\Le
E_{-1/2}\Le \frac{\tilde{U}_{p}}{T_{0}}(k+1)\Ra\,-\,\Le 1\,+\,\frac{m}{\tilde{U}_{p}}\Ra^{3/2}\,
E_{-1/2}\Le \frac{U_{p}}{T_{0}}(k+1)\Ra\Ra\,,
\label{A3}
\eeqar
here $E_{\alpha}(z)$ is the exponential integral function.
Considering \eq{Cru7} limit when
\beq\label{A4}
\tilde{U}_{p}/T_{0}\,,\,\,m/T_{0}\,\gg\,1
\eeq
and expanding the answer with respect to large $\tilde{U}_{p}/T_{0}$, the leading order contributions into the integral we have are the following:
\beq\label{A5}
I\,\approx\,\sum_{k=0}^{\infty}\,\Le \frac{1}{k+1}\,\frac{T_{0}}{\tilde{U}_{p}}\,+\,\frac{1}{2}\frac{1}{(k+1)^{2}}\Le \frac{T_{0}}{\tilde{U}_{p}}\Ra^{2}\,
-\,\frac{e^{-\frac{m}{T_{0}}(k+1)}}{k+1}\,\sqrt{1\,+\,\frac{m}{\tilde{U}_{p}}}\,\frac{T_{0}}{\tilde{U}_{p}}\,-\,\frac{1}{2}\,
\frac{e^{-\frac{m}{T_{0}}(k+1)}}{\Le k+1\Ra^{2}}\,\frac{1}{\sqrt{1\,+\,\frac{m}{\tilde{U}_{p}}}}\,\Le\frac{T_{0}}{\tilde{U}_{p}}\Ra^{2}
\Ra\,.
\eeq
The first term of the expression is divergent. The reason for the divergence is simple, the lowest possible energy for the particle we consider is zero,
that is not quite correct for non-zero temperatures in the classical case and not correct at all for the quantum systems. So, we regularize the lower limit of the integral introducing 
$\mathcal{E}_{0}$ as a lowest non-zero energy for the particle. We have in this case for the \eq{A3} integral
\beqar
&\,&\, I\,=\,\int\limits^{m/\tilde{U}_{p}}_{\mathcal{E}_{0}/\tilde{U}_{p}}\,\frac{\Le z\,+\,1\Ra^{1/2}}{e^{\tilde{U}_{p}(r_{cr})z/T_{0}}\,-\,1}\,dz\,=\,
\sum_{k=0}^{\infty}\int\limits^{m/\tilde{U}_{p}}_{\mathcal{E}_{0}/\tilde{U}_{p}}\,\Le z\,+\,1\Ra^{1/2}\,e^{-(k+1)\tilde{U}_{p}(r_{cr})z/T_{0}}\,dz\,=\,
\nonumber \\
&=&
\sum_{k=0}^{\infty} e^{\frac{\tilde{U}_{p}}{T_{0}}(k+1)}\,\Le (1+\mathcal{E}_{0}/\tilde{U}_{p})^{3/2}
E_{-1/2}\Le \frac{\tilde{U}_{p}}{T_{0}}\,(1+\mathcal{E}_{0}/\tilde{U}_{p} )(k+1)\Ra-\Le 1+\frac{m}{\tilde{U}_{p}}\Ra^{3/2}
E_{-1/2}\Le \frac{U_{p}}{T_{0}}(k+1)\Ra\Ra\, ,
\label{A3001}
\eeqar
that provides:
\beqar
I\,&\approx&\,\sum_{k=0}^{\infty}\,\Le 
\,\frac{e^{-\frac{\mathcal{E}_{0}}{T_{0}}(k+1)}}{k+1}\,\sqrt{1\,+\,\frac{\mathcal{E}_{0}}{\tilde{U}_{p}}}\,\frac{T_{0}}{\tilde{U}_{p}}\,+\,
\frac{1}{2}\,\frac{e^{-\frac{\mathcal{E}_{0}}{T_{0}}(k+1)}}{\Le k+1\Ra^{2}}\,\frac{1}{\sqrt{1\,+\,\frac{\mathcal{E}_{0}}{\tilde{U}_{p}}}}\,
\Le\frac{T_{0}}{\tilde{U}_{p}}\Ra^{2}-\,
\frac{e^{-\frac{m}{T_{0}}(k+1)}}{k+1}\,\sqrt{1\,+\,\frac{m}{\tilde{U}_{p}}}\,\frac{T_{0}}{\tilde{U}_{p}}\,-\,
\right.
\nonumber \\
&-&
\left.
\frac{1}{2}\,
\frac{e^{-\frac{m}{T_{0}}(k+1)}}{\Le k+1\Ra^{2}}\,\frac{1}{\sqrt{1\,+\,\frac{m}{\tilde{U}_{p}}}}\,\Le\frac{T_{0}}{\tilde{U}_{p}}\Ra^{2}
\Ra\,.
\label{A3002}
\eeqar
Performing the summation we obtain finally:
\beqar
I\,&=&\,-\,\ln\Le 1\,-\,e^{-\,\mathcal{E}_{0}/T_{0}}\Ra\,\sqrt{1\,+\,\frac{\mathcal{E}_{0}}{\tilde{U}_{p}}}\,\frac{T_{0}}{\tilde{U}_{p}}\,+\,
\frac{1}{2\sqrt{1\,+\,\frac{\mathcal{E}_{0}}{\tilde{U}_{p}}}}\,Li_{2}(e^{-\frac{\mathcal{E}_{0}}{T_{0}}})\,\Le \frac{T_{0}}{\tilde{U}_{p}}\Ra^{2}\,+\,
\ln\Le 1\,-\,e^{-\frac{m}{T_{0}}}\Ra\,\frac{T_{0}}{\tilde{U}_{p}}\,\sqrt{1\,+\,\frac{m}{\tilde{U}_{p}}}\,-\,
\nonumber \\
&-&
\frac{1}{2\sqrt{1\,+\,\frac{m}{\tilde{U}_{p}}}}\,Li_{2}(e^{-\frac{m}{T_{0}}})\,\Le \frac{T_{0}}{\tilde{U}_{p}}\Ra^{2}\,
\label{A300301}
\eeqar
with $Li_{2}(z)$ as a polylogarithm function.
Therefore, in this approximation, we obtain the following answer for the \eq{A1} expression:
\beqar
N_{1}\,&=&\,\frac{2\pi }{h^{3}}\,V \Le 2 m \tilde{U}_{p}\Ra^{3/2}\,\frac{T_{0}}{\tilde{U}_{p}}\,\Le\,
-\,\ln\Le 1\,-\,e^{-\,\mathcal{E}_{0}/T_{0}}\Ra\,\sqrt{1\,+\,\frac{\mathcal{E}_{0}}{\tilde{U}_{p}}}\,+\,
\frac{1}{2\sqrt{1\,+\,\frac{\mathcal{E}_{0}}{\tilde{U}_{p}}}}\,Li_{2}(e^{-\frac{\mathcal{E}_{0}}{T_{0}}})\,\frac{T_{0}}{\tilde{U}_{p}}\,+\,
\right.
\nonumber \\
&+&
\left.
\ln\Le 1\,-\,e^{-\frac{m}{T_{0}}}\Ra\,\sqrt{1\,+\,\frac{m}{\tilde{U}_{p}}}\,-\,
\frac{1}{2\sqrt{1\,+\,\frac{m}{\tilde{U}_{p}}}}\,Li_{2}(e^{-\frac{m}{T_{0}}})\,\frac{T_{0}}{\tilde{U}_{p}}\Ra\,.
\label{A3003}
\eeqar
In the limit when 
\beq\label{A3004}
\mathcal{E}_{0}\sim\,T_{min}\,\ll\,T_{0}\,,\,\,\,\,\,m/T_{0}\,\gg\,1\,,
\eeq
we have correspondingly for the leading contribution
\beq\label{A300302}
N_{1}\,=\,\frac{2\pi }{h^{3}}\,V \Le 2 m \tilde{U}_{p}\Ra^{3/2}\,\frac{T_{0}}{\tilde{U}_{p}}\,\sqrt{1\,+\,\frac{\mathcal{E}_{0}}{\tilde{U}_{p}}}\,
\Le\,\ln\Le T_{0}/\mathcal{E}_{0}\Ra\,+\,
\frac{\pi^2}{12}\,\frac{T_{0}}{\tilde{U}_{p}\,+\,\mathcal{E}_{0}}\Ra\,\approx\,\frac{2\pi }{h^{3}}\,V \Le 2 m \tilde{U}_{p}\Ra^{3/2}\,\frac{T_{0}}{\tilde{U}_{p}}\,\ln\Le T_{0}/\mathcal{E}_{0}\Ra\,.
\eeq
The similar answer we obtain in the case of the \eq{Cru10} limit, , we need to account only that
\beq\label{A1302}
m/T_{0}\,<\,1\,,
\eeq
obtaining to the leading order with respect to $\frac{T_{0}}{\tilde{U}_{p}}$
\beq\label{A14}
N_{1}\,=\,
\frac{2\pi }{h^{3}}\,V \Le 2 m \tilde{U}_{p}\Ra^{3/2}\,\frac{T_{0}}{\tilde{U}_{p}}\,\Le\,
\ln\Le T_{0}/\mathcal{E}_{0}\Ra\,\sqrt{1\,+\,\frac{\mathcal{E}_{0}}{\tilde{U}_{p}}}\,-\,
\ln\Le T_{0}/m \Ra\,\sqrt{1\,+\,\frac{m}{\tilde{U}_{p}}}\Ra\,\approx\,
\frac{2\pi }{h^{3}}\,V \Le 2 m \tilde{U}_{p}\Ra^{3/2}\,\frac{T_{0}}{\tilde{U}_{p}}\,
\ln\Le m/\mathcal{E}_{0}\Ra\,.
\eeq
Indeed, the assumptions behind the \eq{A5} and \eq{A3002} derivation are still valid and the answer is not changing drastically but only corrected.

 {The \eq{Cru1002} limit, in turn, requires an use of the approximation different from the used in \eq{A3} answer valid
in the case of the \eq{Cru7} limit.} In this case it is suitable we rewrite the \eq{A2} integral in the following form:
\beq\label{A15}
I\,=\,\Le \frac{T_{0}}{\tilde{U}_{p}}\Ra^{3/2}\,\sum_{k=0}^{\infty}\,\Le k\,+\,1 \Ra^{-3/2}\,
\int\limits^{m (k+1)/T_{0}}_{\mathcal{E}_{0} (k+1)/T_{0}}\,\Le z\,+\,\frac{\tilde{U}_{p}}{T_{0}}(k+1)\Ra^{1/2}\,e^{- z}\,dz\,
\eeq
which is fully equivalent to \eq{A3} of course. The calculation of the integral provides:
\beq\label{A16}
I\,=\,\Le \frac{T_{0}}{\tilde{U}_{p}}\Ra^{3/2}\,\sum_{k=0}^{\infty}\,\Le k\,+\,1 \Ra^{-3/2}\,e^{\frac{\tilde{U}_{p}}{T_{0}}\,(k+1)}\,\Le
\Gamma[3/2,\Le \mathcal{E}_{0}+\tilde{U}_{p}\Ra\,(k+1)/T_{0}]\,-\,\Gamma[3/2,U_{p}\,(k+1)/T_{0}]\Ra\,
\eeq
with $\Gamma$ as incomplete gamma function.
The answer is,  {of course, precise, but we are interested in exploring}  the asymptotic behavior of the integral in the limit determined by \eq{Cru1002}.
Direct use of the expansion of the function in respect to the small parameter is problematic now, the summation over $k$ is going to infinity and makes the small parameter in the function large at some $k$.
Therefore, first of all, we separate the full sum on the two parts introducing some parameter $k_{max}\,=\,N$ through the following identity:
\beq\label{A16001}
\Le \mathcal{E}_{0}+\tilde{U}_{p}\Ra\,\Le N\,+\,1\Ra/T_{0}\,=\,\frac{T_{0}}{\tilde{U}_{p}}\,,\,\,\,N\,+\,1\,\approx\,\textrm{floor}[\Le T_{0}/\tilde{U}_{p}\Ra^{2}]\,;
\,\,\,N\,\gg\,1\,.
\eeq
Thereafter consider two different parts of the \eq{A16} integral:
\beqar
I\,&=&\,I_{1}\,+\,I_{2}\,=\,\Le \frac{T_{0}}{\tilde{U}_{p}}\Ra^{3/2}\,\sum_{k=0}^{N-1}\,\Le k\,+\,1 \Ra^{-3/2}\,e^{\frac{\tilde{U}_{p}}{T_{0}}\,(k+1)}\,\Le
\Gamma[3/2,\Le \mathcal{E}_{0}+\tilde{U}_{p}\Ra\,(k+1)/T_{0}]\,-\,\Gamma[3/2,U_{p}\,(k+1)/T_{0}]\Ra\,+\,
\nonumber \\
&+&\,\Le \frac{T_{0}}{\tilde{U}_{p}}\Ra^{3/2}\,
\sum_{k=N}^{\infty}\,\Le k\,+\,1 \Ra^{-3/2}\,e^{\frac{\tilde{U}_{p}}{T_{0}}\,(k+1)}\,\Le
\Gamma[3/2,\Le \mathcal{E}_{0}+\tilde{U}_{p}\Ra\,(k+1)/T_{0}]\,-\,\Gamma[3/2,U_{p}\,(k+1)/T_{0}]\Ra\,.
\label{A17}
\eeqar
The second term in the sum is equivalent to the \eq{A3002} answer with only difference that the sum over $k$ begins from $N$.
Therefore, after the shift of $k$ 
\beq\label{A18}
s\,=\,k\,-\,N
\eeq
we rewrite the second term of \eq{A17} as:
\beqar\label{A19}
I_{2}\,&=&\,
\sum_{s=0}^{\infty} e^{\frac{\tilde{U}_{p}}{T_{0}}(s+N+1)}\,\Le (1+\mathcal{E}_{0}/\tilde{U}_{p})^{3/2}
E_{-1/2}\Le \frac{\tilde{U}_{p}}{T_{0}}\,(1+\mathcal{E}_{0}/\tilde{U}_{p} )(s+N+1)\Ra-\Le 1+\frac{m}{\tilde{U}_{p}}\Ra^{3/2}
E_{-1/2}\Le \frac{U_{p}}{T_{0}}(s+N+1)\Ra\Ra\,=
\nonumber \\
&=&
\sum_{s=0}^{\infty} e^{\frac{\tilde{U}_{p}}{T_{0}}s+\frac{T_{0}}{\tilde{U}_{p}}}\Le (1+\mathcal{E}_{0}/\tilde{U}_{p})^{3/2}
E_{-1/2}\Le (1+\mathcal{E}_{0}/\tilde{U}_{p} )(\frac{\tilde{U}_{p}}{T_{0}}\,s+\frac{T_{0}}{\tilde{U}_{p}})\Ra-\Le 1+\frac{m}{\tilde{U}_{p}}\Ra^{3/2}
E_{-1/2}\Le \frac{U_{p}}{T_{0}}s+\frac{U_{p}T_{0}}{\tilde{U}_{p}^{2}})\Ra\Ra\,.
\eeqar
The arguments of all functions in the expression are large now and the functions can be expended in respect to $\frac{T_{0}}{\tilde{U}_{p}}$ ratio parameter. 
Preserving the only leading order contributions of
the expansion we obtain:
\beqar
&\,& I_{2}\approx\sum_{k=0}^{\infty}\Le 
e^{-\frac{\mathcal{E}_{0}T_{0}}{\tilde{U}_{p}^{2}}} e^{-\frac{T_{0}}{\tilde{U}_{p}}\Le 1+\frac{\mathcal{E}_{0}}{\tilde{U}_{p}} \Ra k}
\Le \sqrt{1+\frac{\mathcal{E}_{0}}{\tilde{U}_{p}}}\frac{\tilde{U}_{p}}{T_{0}}+
\frac{1}{2\sqrt{1+\frac{\mathcal{E}_{0}}{\tilde{U}_{p}}}}\Le\frac{\tilde{U}_{p}}{T_{0}}\Ra^{2}\Ra+
k e^{-\frac{\mathcal{E}_{0}T_{0}}{\tilde{U}_{p}^{2}}}\,e^{-\frac{T_{0}}{\tilde{U}_{p}}\Le 1+\frac{\mathcal{E}_{0}}{\tilde{U}_{p}} \Ra k}
\sqrt{1+\frac{\mathcal{E}_{0}}{\tilde{U}_{p}}}\Le\frac{\tilde{U}_{p}}{T_{0}}\Ra^{2}-
\right.
\nonumber \\
&-&
\left.
e^{-\frac{m T_{0}}{\tilde{U}_{p}^{2}}} e^{-\frac{T_{0}}{\tilde{U}_{p}}\Le 1+\frac{m}{\tilde{U}_{p}} \Ra k}
\Le \sqrt{1 +\frac{m}{\tilde{U}_{p}}}\frac{\tilde{U}_{p}}{T_{0}} +
\frac{1}{2\sqrt{1+\frac{m}{\tilde{U}_{p}}}}\Le\frac{\tilde{U}_{p}}{T_{0}}\Ra^{2}\Ra +
k e^{-\frac{m T_{0}}{\tilde{U}_{p}^{2}}} e^{-\frac{T_{0}}{\tilde{U}_{p}}
\Le 1+\frac{m}{\tilde{U}_{p}} \Ra k}\sqrt{1+\frac{m}{\tilde{U}_{p}}}\,\Le\frac{\tilde{U}_{p}}{T_{0}}\Ra^{2}\,\Ra\,.
\label{A20}
\eeqar
After the summation, the answer we have is the following one:
\beqar
&\,& I_{2}\,=\,
\frac{e^{T_{0}/\tilde{U}_{p}}\,\sqrt{1+\frac{\mathcal{E}_{0}}{\tilde{U}_{p}}}}{e^{\frac{T_{0}}{\tilde{U}_{p}}\Le 1+\frac{\mathcal{E}_{0}}{\tilde{U}_{p}} \Ra}\,-\,1}
\Le 1\,+\,\frac{1}{2\Le 1\,+\,\frac{\mathcal{E}_{0}}{\tilde{U}_{p}}\Ra}\frac{\tilde{U}_{p}}{T_{0}}\Ra\,\frac{\tilde{U}_{p}}{T_{0}}\,+\,
\frac{e^{T_{0}/\tilde{U}_{p}}\,\sqrt{1+\frac{\mathcal{E}_{0}}{\tilde{U}_{p}}}}{\Le e^{\frac{T_{0}}{\tilde{U}_{p}}\Le 1+\frac{\mathcal{E}_{0}}{\tilde{U}_{p}} \Ra}\,-\,1\Ra^{2}}
\Le \frac{\tilde{U}_{p}}{T_{0}}\Ra^{2}\,-\,
\nonumber \\
&-&
\frac{e^{T_{0}/\tilde{U}_{p}}\,\sqrt{1+\frac{m}{\tilde{U}_{p}}}}{e^{\frac{T_{0}}{\tilde{U}_{p}}\Le 1+\frac{m}{\tilde{U}_{p}} \Ra}\,-\,1}
\Le 1\,+\,\frac{1}{2\Le 1\,+\,\frac{m}{\tilde{U}_{p}}\Ra}\frac{\tilde{U}_{p}}{T_{0}}\Ra\,\frac{\tilde{U}_{p}}{T_{0}}\,+\,
\frac{e^{T_{0}/\tilde{U}_{p}}\,\sqrt{1+\frac{m}{\tilde{U}_{p}}}}{\Le e^{\frac{T_{0}}{\tilde{U}_{p}}\Le 1+\frac{m}{\tilde{U}_{p}} \Ra}\,-\,1\Ra^{2}}
\Le \frac{\tilde{U}_{p}}{T_{0}}\Ra^{2}\,.
\label{A21}
\eeqar
The expression is finite in the $\mathcal{E}_{0}\,\rightarrow\,0$ limit, so taking $\mathcal{E}_{0}\,=\,0$ we obtain in turn:
\beqar
&\,& I_{2}\,=\,
\frac{\tilde{U}_{p}/T_{0}}{1\,-\,e^{-\frac{T_{0}}{\tilde{U}_{p}}}}\,\Le 1\,+\,\frac{1}{2}\,\frac{\tilde{U}_{p}}{T_{0}}\Ra\,+\,
\frac{e^{-T_{0}/\tilde{U}_{p}}}{\Le 1\,-\,e^{-T_{0}/\tilde{U}_{p}}\Ra^{2}}\Le \frac{\tilde{U}_{p}}{T_{0}}\Ra^{2}\,-\,
\nonumber \\
&-&
\frac{e^{-m/\tilde{U}_{p}}\,\sqrt{1+\frac{m}{\tilde{U}_{p}}}}{1\,-\,e^{-\,\frac{T_{0}}{\tilde{U}_{p}}\Le 1+\frac{m}{\tilde{U}_{p}} \Ra}}
\Le 1\,+\,\frac{1}{2\Le 1\,+\,\frac{m}{\tilde{U}_{p}}\Ra}\frac{\tilde{U}_{p}}{T_{0}}\Ra\,\frac{\tilde{U}_{p}}{T_{0}}\,+\,
\frac{e^{-\,2\,m/\tilde{U}_{p}}\,\sqrt{1+\frac{m}{\tilde{U}_{p}}}}{\Le 1\,-\, e^{-\frac{T_{0}}{\tilde{U}_{p}}\Le 1+\frac{m}{\tilde{U}_{p}} \Ra}\Ra^{2}}
\Le \frac{\tilde{U}_{p}}{T_{0}}\Ra^{2}\,.
\label{A22}
\eeqar
The first term in \eq{A17} {cannot be expanded in a similar way}. At small $k$ the function's argument is small but it grows at larger $k$ values.
Therefore, first of all, we need to clarify the behavior of the sum. Let's shift the $k$ index in the second term of $I_{1}$ by
\beq\label{A22001}
s+1\,=\,\frac{U_{P}}{\tilde{U}_{p}}\,(k+1)\,=\,(k+1)\,\Le 1\,-\frac{m}{U_{p}}\Ra^{-1}\,;\,\,\,\frac{m}{U_{p}}\,<\,1\,.
\eeq
Taking $\mathcal{E}_{0}=0$, we obtain correspondingly:
\beqar
I_{1}\,&=&\,\sum_{k=0}^{N-1}\,\Le k\,+\,1 \Ra^{-3/2}\,e^{\frac{\tilde{U}_{p}}{T_{0}}\,(k+1)}\,
\Gamma[3/2,\tilde{U}_{p}\,(k+1)/T_{0}]\,-\,
\nonumber \\
&-&\,\Le \frac{T_{0}}{\tilde{U}_{p}}\Ra^{3/2}\,
\sum_{s=s_{min}}^{s_{max}}\,
\Le s\,+\,1 \Ra^{-3/2}\,
e^{\frac{\tilde{U}_{p}}{T_{0}}\,(s+1)\,(1-m/U_{p})}\,\Gamma[3/2,\tilde{U}_{p}\,(s+1)/T_{0}]\,.
\label{A22002}
\eeqar
with
\beq\label{A22003}
s_{min}\,+\,1\,=\,\textrm{floor}[U_{p}/\tilde{U}_{p}]\,=\,\textrm{floor}[1\,+\,m/\tilde{U}_{p}]\,;\,
\,\,\,s_{max}\,=\,\textrm{floor}[U_{P} N/\tilde{U}_{p}]\,=\,\textrm{floor}[U_{p}T_{0}^{2}/\tilde{U}_{p}^{3}]\,,
\eeq
here the $\Le \frac{T_{0}}{\tilde{U}_{p}}\Ra^{3/2}$ factor in the front of \eq{A15} integral we will account directly in the final expression for the $N_{1}$.
The second term of \eq{A22002} can be approximated by replacing of it with a largest value of the term:
\beqar
 I_{1}\,&\approx&\, 
\sum_{k=0}^{N-1}\,\Le k\,+\,1 \Ra^{-3/2}\,e^{\frac{\tilde{U}_{p}}{T_{0}}\,(k+1)}\,\Gamma[3/2,\tilde{U}_{p}\,(s+1)/T_{0}]\,-\,
\nonumber \\
&-&\,
\Le \frac{U_{p}}{\tilde{U}_{p}}\Ra^{3/2}\,e^{-\frac{m \tilde{U}_{p}}{T_{0}U_{p}}\,(s_{min}+1)}\,\sum_{k=s_{min}}^{s_{max}}\,
\Le k\,+\,1 \Ra^{-3/2}\,e^{\frac{\tilde{U}_{p}}{T_{0}}\,(k+1)}\,\Gamma[3/2,\tilde{U}_{p}\,(k+1)/T_{0}]\,=\,
\nonumber \\
&=&
\,\sum_{k=0}^{N-1}\,\Le k\,+\,1 \Ra^{-3/2}\,e^{\frac{\tilde{U}_{p}}{T_{0}}\,(k+1)}\,
\Gamma[3/2,\tilde{U}_{p}\,(s+1)/T_{0}]\,-\,
\nonumber \\
&-&\,e^{-\frac{m }{T_{0}}}\,\Le \frac{U_{p}}{\tilde{U}_{p}}\Ra^{3/2}\,\sum_{k=s_{min}}^{s_{max}}\,
\Le k\,+\,1 \Ra^{-3/2}\,e^{\frac{\tilde{U}_{p}}{T_{0}}\,(k+1)}\,\Gamma[3/2,\tilde{U}_{p}\,(k+1)/T_{0}]\,.
\label{A22004}
\eeqar
Next we approximate the $I_{1}$ sum by an integration. Namely, let's define
\beq\label{A23}
\Delta\,=\,\frac{\tilde{U}_{p}}{T_{0}}\,\ll\,1\,,\,\,\,x_{k}\,=\,\Delta\Le k\,+\,1\Ra\,;\,\,\,x_{0}\,=\,\Delta\,,\,\,\,
x_{N}\,=\,\frac{T_{0}}{\tilde{U}_{p}}\,=\,1/\Delta\,;\,
\eeq
and
\beq\label{A23001}
x_{min}\,=\,\Delta(s_{min}+1)\,=\,U_{p}/T_{0}\,,\,\,\,x_{max}\,=\,\Delta(s_{max}+1)\,=\,U_{p}\, T_{0}/\tilde{U}_{p}^{2}\,.
\eeq
Then we rewrite the $I_{1}$ sum as an integral:
\beq\label{A24}
I_{1}\,\approx\,\Delta^{1/2}\,\Le \int_{x_{0}}^{x_{N}}\,\frac{\Gamma[3/2,x]}{x^{3/2}}\,e^{x}\,dx\,-\,
e^{-\frac{m }{T_{0}}}\,\Le \frac{U_{p}}{\tilde{U}_{p}}\Ra^{3/2}\,
\int_{x_{min}}^{x_{max}}\,\frac{\Gamma[3/2,U_{p}\, x/\tilde{U}_{p}]}{x^{3/2}}\,e^{x}\,dx \Ra\,.
\eeq
By the use of a series expansion of the incomplete gamma function:
\beq\label{A25}
\Gamma[3/2,x]\,=\,\Gamma[3/2]\Le 1\,-\,x^{3/2}\,e^{-x}\,\sum_{k=0}^{\infty}\,\frac{x^{k}}{\Gamma[5/2+k]}\Ra\,
\eeq
we obtain for the first integral
\beqar
I_{1 a}\,&=&\,\Delta^{1/2}\,\int_{x_{0}}^{x_{N}}\,\frac{\Gamma[3/2,x]}{x^{3/2}}\,e^{x}\,dx\,=\,
\Delta^{1/2}\,\Gamma[3/2]\,\Le \int_{\Delta}^{1/\Delta}\,\frac{e^{x}}{x^{3/2}}\,dx\,-\,
\sum_{k=0}^{\infty}\,\frac{1}{\Gamma[5/2+k]}\,\int_{\Delta}^{1/\Delta}\,x^{k}\,dx\,\Ra\,=\,
\nonumber \\
&=&
\frac{\sqrt{\pi}}{2}\,\Delta^{1/2}\,\Le
\frac{2}{\sqrt{\Delta}}\Le e^{\Delta}\,+\,\Delta\,E_{1/2}(-\Delta)\Ra\,-\,\frac{2}{\sqrt{\Delta}}\,
\Le \Delta\,e^{1/\Delta}\,+\,E_{1/2}(-1/\Delta)\Ra\,-\,
\right.
\nonumber \\
&-&\,
\left.
\frac{4}{3\sqrt{\pi}\Delta}\,\,_{2}F_{2}[(1,1),(2,5/2),1/\Delta]\,+\,\frac{4\Delta}{3\sqrt{\pi}}\,\,_{2}F_{2}[(1,1),(2,5/2),\Delta]\,
\Ra\,.
\label{A26}
\eeqar
The correct asymptotic behavior of the sum can be reconstructed now by $\Delta\rightarrow\,0$ limit taking. Preserving the first two term from the asymptotic expansion of the functions, we have for the first term of \eq{A24}:
\beqar
I_{1 a}\,&\approx &\,\frac{\sqrt{\pi}}{2}\,\Delta^{1/2}\,\Le
\frac{2}{\sqrt{\pi}}\Le \ln(1/\Delta)\,-\,\psi(3/2)\Ra\,+\,
\frac{2}{\sqrt{\Delta}}\,-\,2\sqrt{\Delta}\,+\,\frac{\Delta}{3\sqrt{\pi}}\,-\,
\frac{\Delta^{3/2}}{3}\,+\,\frac{31\Delta^{2}}{60\sqrt{\pi}}\Ra\,=\,
\nonumber \\
&=&
\,\Le\ln(\frac{T_{0}}{\tilde{U}_{p}})\,-\,\psi(3/2)\Ra\,\Le\frac{\tilde{U}_{p}}{T_{0}}\Ra^{1/2}\,+\,\sqrt{\pi}\,-\,\sqrt{\pi}\,\frac{\tilde{U}_{p}}{T_{0}}\,+\,
\frac{1}{6}\,\Le\frac{\tilde{U}_{p}}{T_{0}}\Ra^{3/2}\,-\,\frac{\sqrt{\pi}}{6}\,\Le\frac{\tilde{U}_{p}}{T_{0}}\Ra^{2}\,+\,
\frac{31}{120}\,\Le\frac{\tilde{U}_{p}}{T_{0}}\Ra^{5/2}\,.
\label{A27}
\eeqar
The second integral in \eq{A24} is different from the first one only by the limits of integration, but is suppressed by $e^{-\frac{m }{T_{0}}}$ factor. 
So, preserving only leading terms of the first contribution, we obtain for the integral:
\beq\label{A28}
I_{1}=\sqrt{\pi}+
\Le \ln(\frac{T_{0}}{\tilde{U}_{p}})-\psi(3/2)\Ra\,\Le\frac{\tilde{U}_{p}}{T_{0}}\Ra^{1/2}\,-\sqrt{\pi}\,\frac{\tilde{U}_{p}}{T_{0}}
+\frac{1}{6}\Le\frac{\tilde{U}_{p}}{T_{0}}\Ra^{3/2}\,.
\eeq
Accounting the factor in the front of the \eq{A16} expression, we obtain correspondingly
\beq\label{A29}
I_{1}\,\approx\,\Le \frac{T_{0}}{\tilde{U}_{p}}\Ra^{3/2}\,\Le \sqrt{\pi}+
\Le \ln(\frac{T_{0}}{\tilde{U}_{p}})-\psi(3/2)\Ra\,\Le\frac{\tilde{U}_{p}}{T_{0}}\Ra^{1/2}\,-\sqrt{\pi}\,\frac{\tilde{U}_{p}}{T_{0}}+
\frac{1}{6}\Le\frac{\tilde{U}_{p}}{T_{0}}\Ra^{3/2}\Ra
\eeq
that provides finally to the leading order contribution the following expression:
\beq\label{A30}
N_{1}\,=\,
\frac{2\pi }{h^{3}}\,V \Le 2 m T_{0}\Ra^{3/2}\,\Le \sqrt{\pi}+
\Le \ln(\frac{T_{0}}{\tilde{U}_{p}})-\psi(3/2)\Ra\,\Le\frac{\tilde{U}_{p}}{T_{0}}\Ra^{1/2}\,
\Ra\,.
\eeq

\newpage
\section{ Integrals for thermodynamic quantities of the crust: energy integrals}\label{AppB}
\renewcommand{\theequation}{B.\arabic{equation}}
\setcounter{equation}{0}

 Integral we calculate now is the following one:
\beq\label{B1}
I\,=\,\int\limits^{m/\tilde{U}_{p}}_{\mathcal{E}_{0}/\tilde{U}_{p}}\,\frac{\Le z\,+\,1\Ra^{3/2}}{e^{\tilde{U}_{p}(r_{cr})\,z/T}\,-\,1}\,dz\,=\,
\sum_{k=0}^{\infty}\int\limits^{m/\tilde{U}_{p}}_{\mathcal{E}_{0}/\tilde{U}_{p}}\,\Le z\,+\,1\Ra^{3/2}\,e^{-(k+1)\tilde{U}_{p}(r_{cr})z/T}\,.
\eeq
The calculations are similar to the done in the previous Appendix, so we obtain the following results preserving the only leading contributions in the answer:
\beqar
I\,&\approx&\,\sum_{k=0}^{\infty}\,\Le 
\,\frac{e^{-\frac{\mathcal{E}_{0}}{T}(k+1)}}{k+1}\,\Le 1\,+\,\frac{\mathcal{E}_{0}}{\tilde{U}_{p}}\Ra^{3/2}\,\frac{T}{\tilde{U}_{p}}\,+\,
\frac{3}{2}\,\frac{e^{-\frac{\mathcal{E}_{0}}{T}(k+1)}}{\Le k+1\Ra^{2}}\,\sqrt{1\,+\,\frac{\mathcal{E}_{0}}{\tilde{U}_{p}}}\,
\Le\frac{T}{\tilde{U}_{p}}\Ra^{2}-\,
\frac{e^{-\frac{m}{T}(k+1)}}{k+1}\,\Le 1\,+\,\frac{m}{\tilde{U}_{p}}\Ra^{3/2}\,\frac{T}{\tilde{U}_{p}}\,-\,
\right.
\nonumber \\
&-&
\left.
\frac{3}{2}\,
\frac{e^{-\frac{m}{T}(k+1)}}{\Le k+1\Ra^{2}}\,\sqrt{1\,+\,\frac{m}{\tilde{U}_{p}}}\,\Le\frac{T}{\tilde{U}_{p}}\Ra^{2}
\Ra\,.
\label{B2}
\eeqar
Performing the summation we obtain:
\beqar
I\,&=&\,-\,\ln\Le 1\,-\,e^{-\,\mathcal{E}_{0}/T}\Ra\,\Le 1\,+\,\frac{\mathcal{E}_{0}}{\tilde{U}_{p}}\Ra^{3/2}\,\frac{T}{\tilde{U}_{p}}\,+\,
\frac{3}{2}\,\sqrt{1\,+\,\frac{\mathcal{E}_{0}}{\tilde{U}_{p}}}\,Li_{2}(e^{-\frac{\mathcal{E}_{0}}{T}})\,\Le \frac{T}{\tilde{U}_{p}}\Ra^{2}\,+\,
\ln\Le 1\,-\,e^{-\frac{m}{T}}\Ra\,\frac{T}{\tilde{U}_{p}}\,\Le 1\,+\,\frac{m}{\tilde{U}_{p}}\Ra^{3/2}\,-\,
\nonumber \\
&-&
\frac{3}{2}\,\sqrt{1\,+\,\frac{m}{\tilde{U}_{p}}}\,Li_{2}(e^{-\frac{m}{T_{0}}})\,\Le \frac{T_{0}}{\tilde{U}_{p}}\Ra^{2}\,.
\label{B3}
\eeqar
Taking into account together the both terms of the \eq{Cru15}, we finally derive the following approximate leading order expression for the internal energy:
\beq
U\,\approx\,\frac{2\pi }{h^{3}}\,V \Le 2 m \tilde{U}_{p}\Ra^{3/2}\,\frac{T}{\tilde{U}_{p}}\,\tilde{U}_{p}\,
\Le\,-\,\ln\Le 1\,-\,e^{-\,\mathcal{E}_{0}/T}\Ra\,\frac{\mathcal{E}_{0}}{\tilde{U}_{p}}\,\sqrt{1\,+\,\frac{\mathcal{E}_{0}}{\tilde{U}_{p}}}\,+\,
\ln\Le 1\,-\,e^{-\frac{m}{T}}\Ra\,\frac{m}{\tilde{U}_{p}}\,\sqrt{1\,+\,\frac{m}{\tilde{U}_{p}}}\Ra\,.
\label{B4}
\eeq
Correspondingly, in the limit 
\beq\label{B5}
\mathcal{E}_{0}\sim\,T_{min}\,\ll\,T\,,\,\,\,\,\,m/T\,\gg\,1\,,
\eeq
we have correspondingly for the leading contribution
\beq\label{B6}
U\,\approx\,\mathcal{E}_{0}\,\frac{2\pi }{h^{3}}\,V\, \Le 2 m \tilde{U}_{p}\Ra^{3/2}\,\frac{T}{\tilde{U}_{p}}\,
\Le \ln\Le T/\mathcal{E}_{0}\Ra\,-\,m\,e^{-\frac{m}{T}}\Ra\,=\,N_{1}\,\Le\mathcal{E}_{0}\,-\,\frac{m e^{-\frac{m}{T}}}{\ln\Le T/\mathcal{E}_{0}\Ra}\Ra\,;
\,\,\,\,U/N_{1}\,=\,\Le\mathcal{E}_{0}\,-\,\frac{m e^{-\frac{m}{T}}}{\ln\Le T/\mathcal{E}_{0}\Ra}\Ra\,;
\eeq
with $N_{1}$ given by \eq{A3003}. For the
\beq\label{B7}
m\,\gg\,\mathcal{E}_{0}\,;\,\,\,\,m/T\,<\,1\,,
\eeq
conditions, we have to the leading order precision:
\beq\label{B8}
U\,\approx\,\frac{2\pi }{h^{3}}\,V\, \Le 2 m \tilde{U}_{p}\Ra^{3/2}\,\frac{T}{\tilde{U}_{p}}\,\Le
m\,\ln\Le m/\mathcal{E}_{0}\Ra\,+\Le \mathcal{E}_{0}\,-\,m \Ra\,\ln\Le T/\mathcal{E}_{0}\Ra\Ra\,
\approx\,
-\,N_{1}\,m\,\frac{\ln\Le T/m \Ra}{\ln\Le m/\mathcal{E}_{0}\Ra}\,,
\eeq
with
\beq\label{B801}
U/N_{1}\,=\,-\,m\,\frac{\ln\Le T/m \Ra}{\ln\Le m/\mathcal{E}_{0}\Ra}\,;
\eeq
here $N_{1}$ is given by \eq{A14}.

 In the \eq{Cru1002} limit, we again rewrite the integral in the following form:
\beq\label{B9}
I\,=\,\Le \frac{T}{\tilde{U}_{p}}\Ra^{5/2}\,\sum_{k=0}^{\infty}\,\Le k\,+\,1 \Ra^{-5/2}\,
\int\limits^{m (k+1)/T_{0}}_{\mathcal{E}_{0} (k+1)/T_{0}}\,\Le z\,+\,\frac{\tilde{U}_{p}}{T}(k+1)\Ra^{3/2}\,e^{- z}\,dz\,
\eeq
which is fully equivalent to \eq{A3} of course. The calculation of the integral provides:
\beq\label{B10}
I\,=\,\sum_{k=0}^{\infty}\,\Le k\,+\,1 \Ra^{-5/2}\,e^{\frac{\tilde{U}_{p}}{T_{0}}\,(k+1)}\,\Le
\Gamma[5/2,\Le \mathcal{E}_{0}+\tilde{U}_{p}\Ra\,(k+1)/T_{0}]\,-\,\Gamma[5/2,U_{p}\,(k+1)/T_{0}]\Ra\,,
\eeq
the $\Le \frac{T_{0}}{\tilde{U}_{p}}\Ra^{5/2}$ factor again we will account in the final expression.
Performing the same approximation as in the previous Section, we arrive to integral we need to calculate:
\beq\label{B11}
I_{1 a}\,=\,\Delta^{3/2}\,\int_{x_{0}}^{x_{N}}\,\frac{\Gamma[5/2,x]}{x^{5/2}}\,e^{x}\,dx\,=\,
\Delta^{3/2}\,\Gamma[5/2]\,\Le \int_{\Delta}^{1/\Delta}\,\frac{e^{x}}{x^{5/2}}\,dx\,-\,
\sum_{k=0}^{\infty}\,\frac{1}{\Gamma[7/2+k]}\,\int_{\Delta}^{1/\Delta}\,x^{k}\,dx\,\Ra\,.
\eeq
Preserving the leading terms in the answer we obtain:
\beqar
I_{1 a}\,&\approx&\,\frac{3\sqrt{\pi}}{4}\Delta^{3/2}\Le
\frac{4}{3\sqrt{\pi}}\Le \ln(1/\Delta)\,-\,\psi(3/2)\Ra\,+\,\frac{2}{3\Delta^{3/2}}\,+\,
\frac{2}{\sqrt{\Delta}}\,-\,\sqrt{\Delta}\,-\,\frac{22\Delta}{15\sqrt{\pi}}\Ra\,=\,
\nonumber \\
&=&\,
\frac{\sqrt{\pi}}{2}\,+\,\Delta^{3/2}\Le \ln(1/\Delta)\,-\,\psi(3/2)\Ra\,+\,\frac{3\sqrt{\pi}}{2}\Delta\,-\,\frac{3\sqrt{\pi}}{4}\Delta^{2}
-\,\frac{11}{2}\Delta^{5/2}\,.
\label{B12}
\eeqar
Preserving the on;y leading terms, we obtain, therefore, for the internal energy in this approximation:
\beq\label{B13}
U\,\approx\,\frac{2\pi }{h^{3}}\,V \Le 2 m T\Ra^{3/2}\,T\,\Le
\frac{\sqrt{\pi}}{2}\,+\,\Le\frac{\tilde{U}_{p}}{T}\Ra^{3/2}\Le \ln(T/\tilde{U}_{p})\,-\,\psi(3/2)\Ra\,
\Ra\,-\,\tilde{U}_{p}\,N_{1}
\eeq
that provides in the leading order approximation:
\beq\label{B14}
U\,\approx\,N_{1}\,T\,\Le 1\,-\,2\,\frac{\tilde{U}_{p}}{T}\Ra/2\,;\,\,\,\,U/N_{1}\,\approx\,T\Le 1/2\,-\,\tilde{U}_{p}/T\Ra\,,
\eeq
see \eq{A30} answer.


\newpage


\bibliographystyle{unsrt}  
\bibliography{core-crust}

\end{document}